# Blockchain-Enhanced Offloading in Mobile Edge Computing: A Systematic Review and Survey of Current Trends and Future Directions

Komeil Moghaddasi[1], Shakiba Rajabi[2]
[1,2]Department of Computer Engineering, Urmia Branch, Islamic Azad University, Urmia, Iran
K.moghaddasi@ieee.org[1], Shakiba.rajabi@ieee.org[2]
Corresponding Author: Komeil Moghaddasi

**Abstract:** With the rapid growth of Internet of Things (IoT) applications, there's a big demand for more processing power and resources in devices. Mobile Edge Computing (MEC) looks promising for enhancing performance and reducing costs by offloading the computing work of IoT to MEC servers. However, the current methods for offloading have issues with privacy and security during the transfer of data and programs. To tackle this, recent developments have introduced secure offloading methods using Blockchain technology, which helps to make MEC more secure by building trust between nodes, improving how edges are authenticated and accessed, and stopping unauthorized access to devices. This paper reviews these Blockchain-based offloading methods for different MEC settings. It starts by explaining the key ideas in offloading and Blockchain, then it sorts the Blockchain-based offloading methods by the algorithms they use. It also compares the offloading methods in each group and ends with a discussion and comparison of the different techniques, tools, and metrics used in these methods.

## 1. Introduction

Internet of Things (IoT) refers to a system that connects a large number of sensors as well as various electronic devices and objects that use wireless links [1] . In addition, the IoT objects can collect some data from their environment or their status and send them occasionally over a network to the remote cloud computing data center [2, 3] .

The edge computing approach brings data processing as close as possible to IoT devices [4]. Additionally, these devices send massive amounts of data at the edge of the network, and therefore the traditional cloud computing paradigm is facing serious bandwidth constraints created by data transmissions through terminal devices [5].

Moreover, cloud computing cannot support ultra-low-latency applications with a reliable Quality of Service (QoS) [6]. Mobile applications that require heavy computation and strict latency requirements are increasingly common on IoT devices with the proliferation of compute-intensive and delay-sensitive applications. Offloading complex computing tasks to mobile cloud computing (MCC) or MEC servers is one promising solution for IoT devices [7]. MEC servers provide lower latency due to their proximity to IoT devices, whereas MCC servers can support complex applications with flexibility and scalability [8, 9]. MEC is an evolving computing paradigm that sinks cloud computing power to the edge, driving innovation in ultra-low latency communication technology and building an intelligent edge platform [10]. Close proximity to the location is the primary benefit of the MEC over other technologies, which can be utilized to enhance real-time immersive interactions and experiences related to ubiquitous intelligent services (e.g., digital twins, smart cities, metaverses, online games, etc) [11, 12], Basically, MEC is a network model that enables cloud-based computation and is the foundation for information technology services at the edge of the cellular network [13]. Basically, MEC's principle is to reduce network traffic and improve application performance by hosting applications nearer to cellular users [14, 15].

To overcome the limitations of smartphones and provide effective computation, offloading techniques are becoming increasingly popular [16, 17].With offloading, users can minimize the costs associated with using cellular networks by transferring as much traffic over WiFi connections as possible [18]. Consequently, the SMD's battery power will be preserved, and it will be able to be utilized for longer periods of time since it can be released for offloaded processing [19]. Moreover, applications are also able to benefit from the powerful processing resources of the remote servers by offloading their processing tasks there. In addition, SMDs now have a higher utility by enabling remote execution, resulting in a wider range of applications being supported [20]. Basically, by offloading, MEC performs significantly better while consuming less energy and latency. In the same way, MEC offloading can facilitate the computation requirements of the end-user devices by bringing the computing nodes to the edge of the network [21]. The offloading decision is necessary due to limited resource availability at the edge of the network and users' favorite computing applications that prefer to consume minimal energy and process data quickly[22].

Obviously, security and privacy concerns have to be taken into consideration as well. There are several methods that can be used to address privacy issues, including the use of blockchain technology. Blockchain technology involves a digital data structure, a shared and distributed database that records transactions in chronological order and maintains a continuously growing log of transactions over time. In other words, a data structure consists of data records, digital transactions, and executables malicious attacks may be conducted against individuals based on IoT data leaks [23]. However, blockchain is a viable option for MECs due to its potential. A win-win solution can be achieved by integrating MEC and blockchain. One of the benefits of blockchain technology for MEC is the privacy and security of data [24]. In addition, MEC can improve the efficiency and scalability of blockchain [25, 26] . Moreover, blockchains are secure because they rely on a proof-of-work algorithm, called mining. Mining is a mathematical problem that is extremely tough to solve, making blockchains almost impossible to hack [27]. A blockchain-based environment for offloading mining tasks is shown in Fig. 1. A blockchain network deploys and manages a three-tier hierarchy consisting of IoT devices, MCC servers, and MEC servers.
The structure of this article is laid out as follows: The second part delves into our research methodology, detailing the process of selecting pertinent articles and formulating research questions to deeply investigate how blockchain technology can be integrated into MEC. The third section provides an in-depth look at blockchain technology and its potential applications. The focus shifts to offloading strategies in the fourth section. In the fifth section, we explore various methods for incorporating blockchain into MEC environments. Practical considerations for selecting features in MEC and blockchain contexts are discussed in the seventh section. The eighth section addresses the challenges and envisages future prospects for MEC. The article is wrapped up in the ninth section, which summarizes the key benefits and challenges associated with melding blockchain technology with MEC.

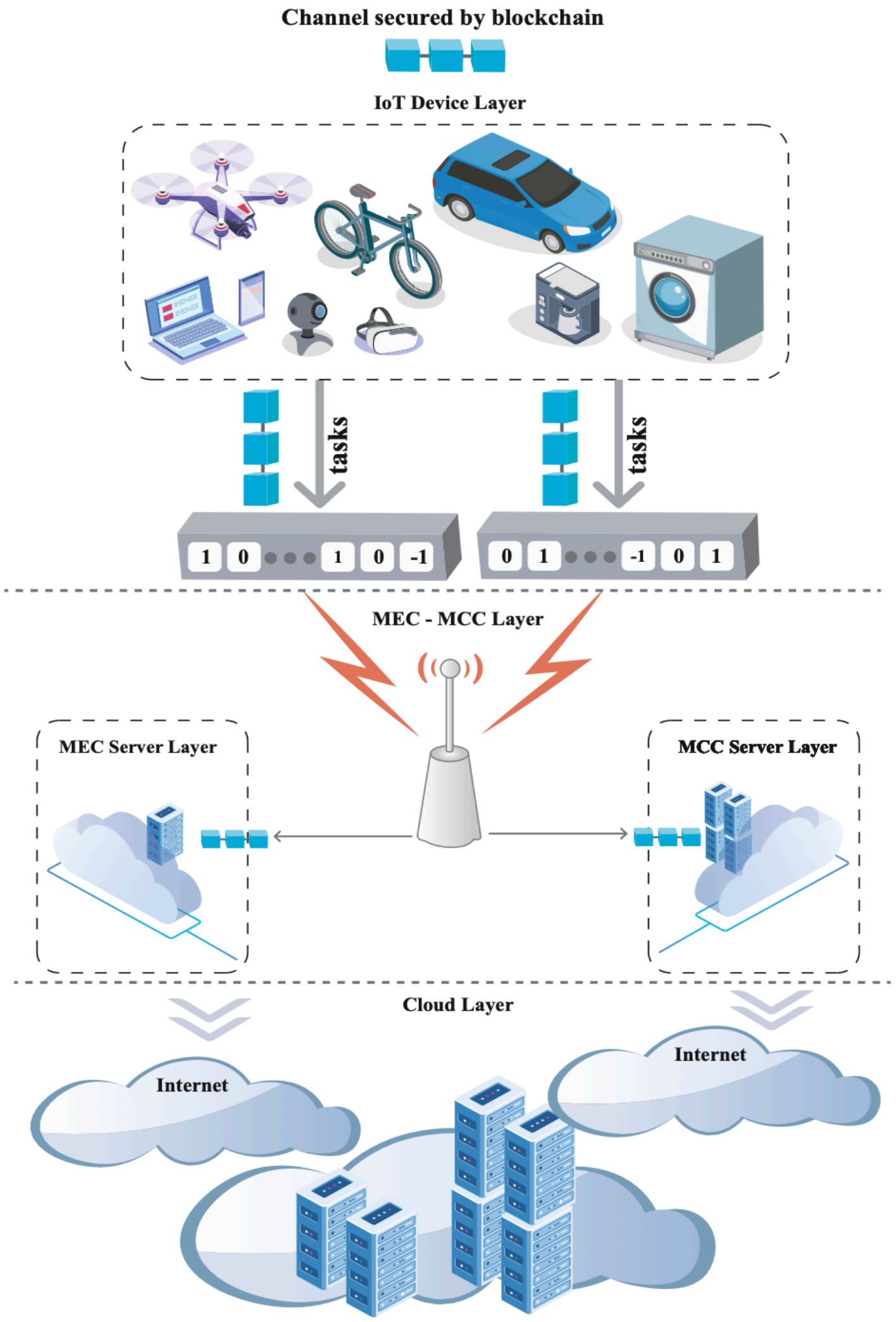

**Fig. 1.** Blockchain-based task offloading framework in the IoT-edge-cloud computing environment

## 2. Research methodology

The purpose of this section is to review recent papers on computation offloading in MEC. Afterward, each survey article will be evaluated for specifications and weaknesses. In the beginning, the following search strings in Google Scholar were used to find reviews and surveys in the stochastic offloading context:

- Offloading survey Mobile edge computing
- Offloading review MEC
- MEC Offloading survey
- Blockchain MEC Offloading survey

**Table 1.** Applied abbreviations

| **Abbreviations** | **Description** |
|---|---|
| MEC | mobile edge computing |
| IoT | Internet of things |
| QoS | Quality of service |
| MCC | Mobile cloud computing |
| SHA | Secure Hash Algorithm |
| CPU | Central Processing Unit |
| ES | Edge servers |
| SMD | Steered Molecular Dynamics |
| NSGA-III | Non-dominated Sorting Genetic Algorithm, the Third Version |
| AIoT | Ambient intelligence is introduced |
| RL | Reinforcement Learning |
| MDP | markov decision process |
| DRL | Deep reinforcement learning |
| DQN | Deep Q-network |
| D3QN | Deep double Q network |
| DDPG | Deep Deterministic Policy Gradient |
| A3C | Asynchronous advantage actor critic |
| DL | Deep learning |
| HRL | hierarchical reinforcement learning |
| HDL | hierarchical deep learning |
| DQL | Deep q-learning |
| DNN | Deep neural network |
| FL | Federated learning |
| AGD | Asynchronous gradient descent |

**Table 2.** Existing surveys and reviews

| **Ref** | **Type of review** | **Year** | **Security** | **Future works and open issues** | **Weakness** |
|---|---|---|---|---|---|
| [28] | Review | 2020 | No | Yes | This survey reviews the articles published earlier than 2020 and is only focused on game-theoretic schemes. |
| [29] | Survey | 2020 | No | Yes | A great number of newly published papers in the field of MEC and offloading have not been covered in this survey. |
| [30] | Survey | 2021 | Yes | No | The absence of a systematic format for selecting the papers. A great number of key articles published after 2019 have not been covered by the present investigation. |
| [31] | Survey | 2022 | No | No | Inadequate directions for future investigation, the unavailability of a systematic format for the purpose of selecting the papers. |
| [32] | Survey | 2022 | Yes | No | Poor categorization of subjects, inadequate robust papers. |

The initial motivation for preparing a survey paper on this topic was to address all these deficiencies. To identify relevant literature, a search strategy was developed for this systematic research. Accordingly, in the search strategy, two databases were targeted: Scopus and Web of Science, and the following search terms were utilized:

- Blockchain AND Mobile Edge Computing
- Blockchain AND MEC
- Blockchain AND Edge AND Offloading
- Blockchain AND MEC AND Offloading
- Blockchain AND Mobile Edge Computing AND Offloading

A variety of sources were searched, including databases and articles published in English only, including journal articles, conference papers, review papers, and case studies.

In the mapping of existing literature Blockchain, MEC, and offloading were the main search terms. In the search, all papers before 2018 were excluded, and the search scope was from 2018 to 2022. During this step, 814 records were extracted from the targeted databases. In order to conduct this study, original research articles, review papers, as well as conference papers were used. It was important that all duplications were thoroughly checked in order to maintain quality in the review. In order to ensure the relevance and quality of the academic literature considered in the review process, the abstracts of the papers were thoroughly reviewed through an analysis and purification process. Each research paper was carefully evaluated at a later stage of the research process to make sure it adhered to the standards that were required. Following this, the next exclusion criterion was the limitation of papers to only those written in the English language. There were four articles in non-English language and they were excluded from the study.

In addition, after the duplicate records are filtered from the study, there are a total of 263 articles that are removed from the analysis. There were also 22 articles that were excluded from the study for other reasons as well. Finally, a total of 50 articles were selected after each article was assessed based on the criteria for inclusion and exclusion mentioned above.

Fig. 2 depicts the inclusion and exclusion of literature at each stage. Moreover, Table 3 shows the distribution of search term results returned by the database.

In the data extraction phase, 50 selected articles were examined that had the following characteristics:

1. Original papers, review articles, and conference papers are included. Published reports and case studies were excluded.
2. Articles must be written in English and be related to the fields of computer science, engineering, or mathematics.
3. The collection of articles was extracted from publications between 2018 and 2022.

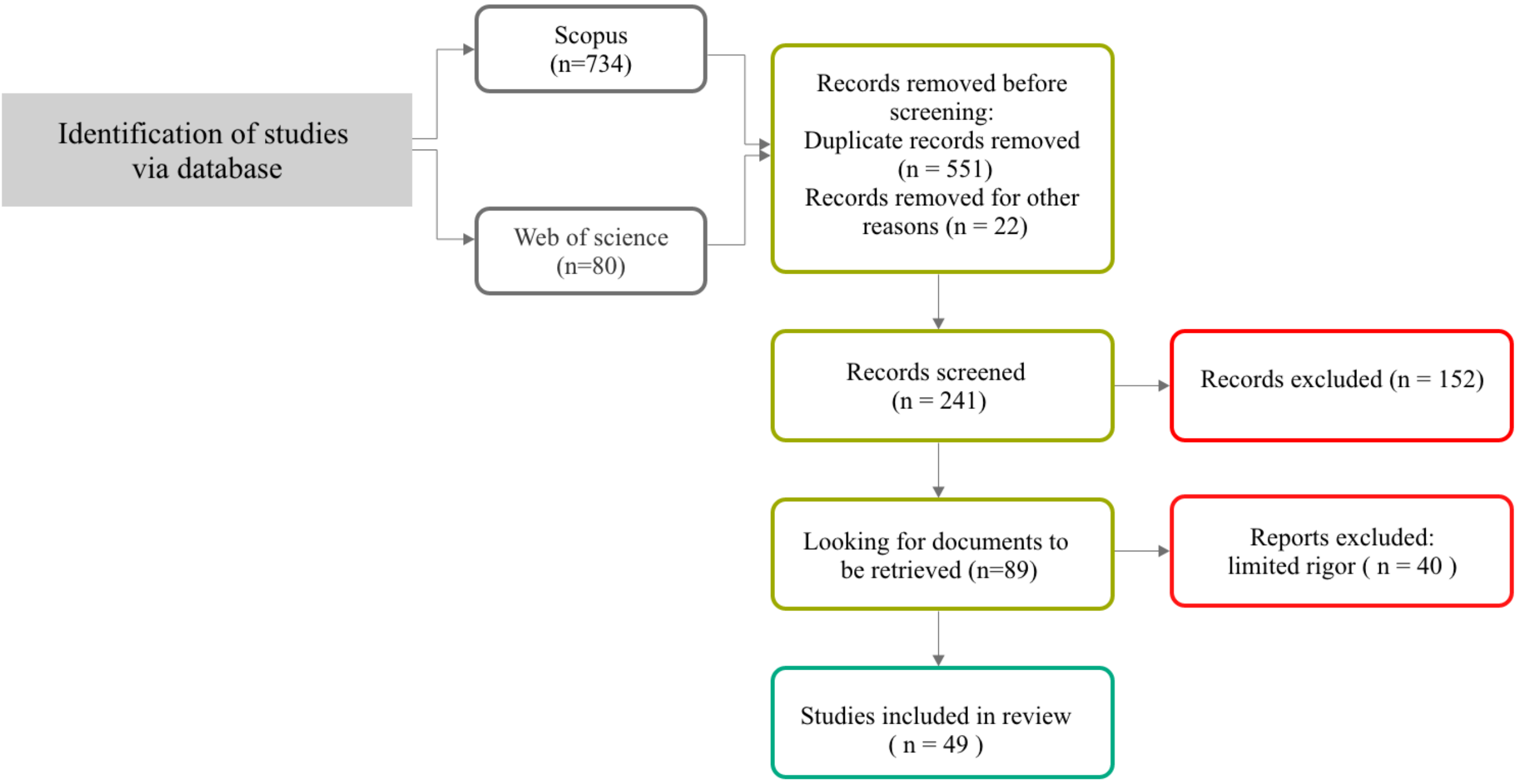

**Fig. 2.** Review process.

**Table 3.** Search terms and results by scholarly databases.

| Search terms vs database (only in Title and Abstract) | Scopus | Web of science |
|---|---|---|
| Blockchain AND "mobile edge computing" | 208 | 37 |
| Blockchain AND MEC | 205 | 12 |
| Blockchain AND edge AND offloading | 170 | 22 |
| Blockchain AND MEC AND offloading | 76 | 1 |
| Blockchain AND "mobile edge computing" AND offloading | 75 | 8 |

### 2.1. Characteristics of included studies

The review is based on a search of publications between 2018 and 2022. Nevertheless, no satisfactory publications were found in all databases that were searched before 2018 that related to this study's scope. From 2018 to 2022, the number of articles grew exponentially, as shown in Fig. 3. A screening process is performed to identify credible original articles, conference papers, and review papers from these searches. For example, the number of journal articles and conference papers is presented in Fig. 4. Papers that did not come from the publishers shown in Fig. 5 were excluded. As for the remaining articles, they were from journals and conferences in Fig 6, which were used in the research and will be reviewed in the following sections.

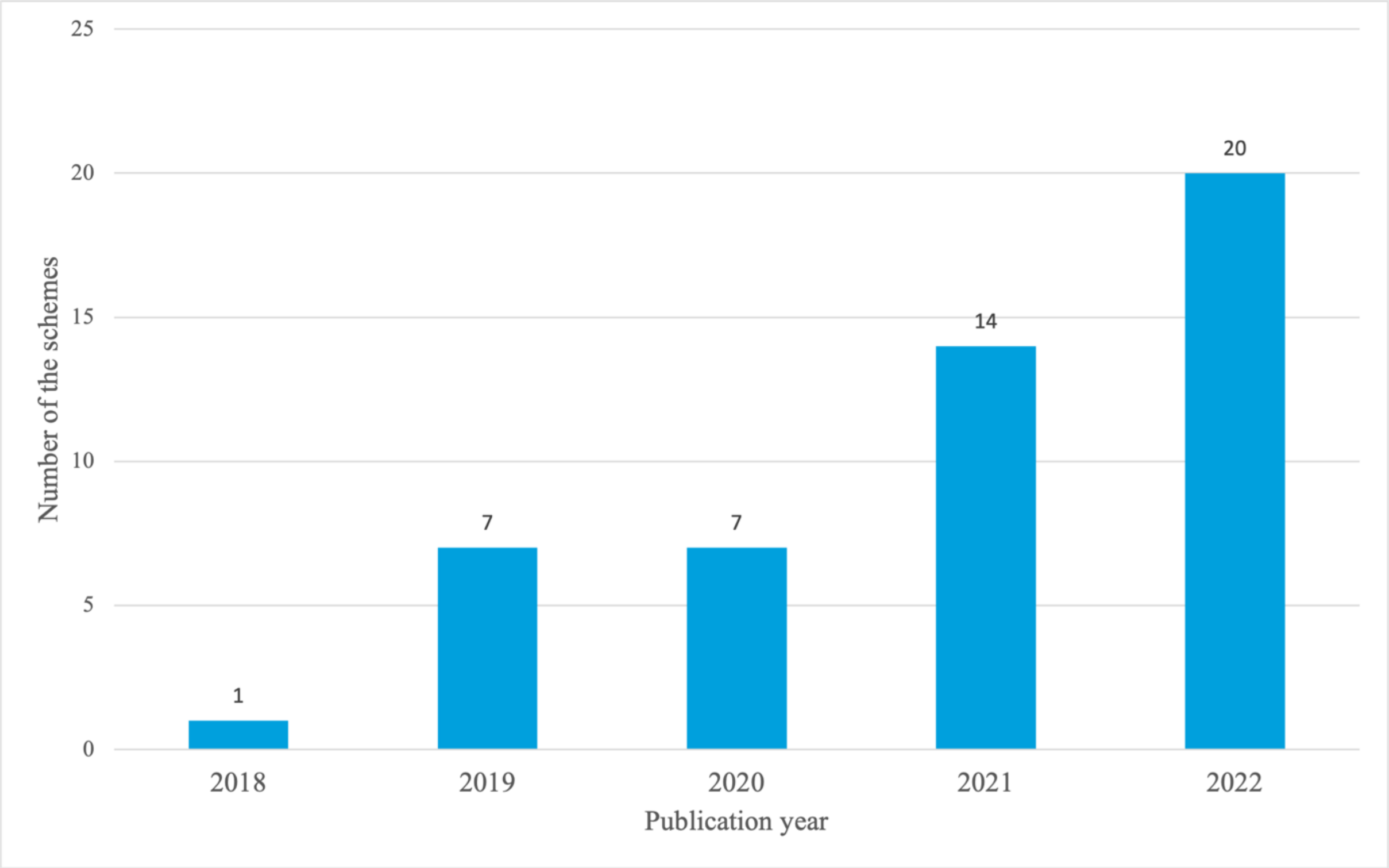

**Fig. 3**. Number of articles published by each year

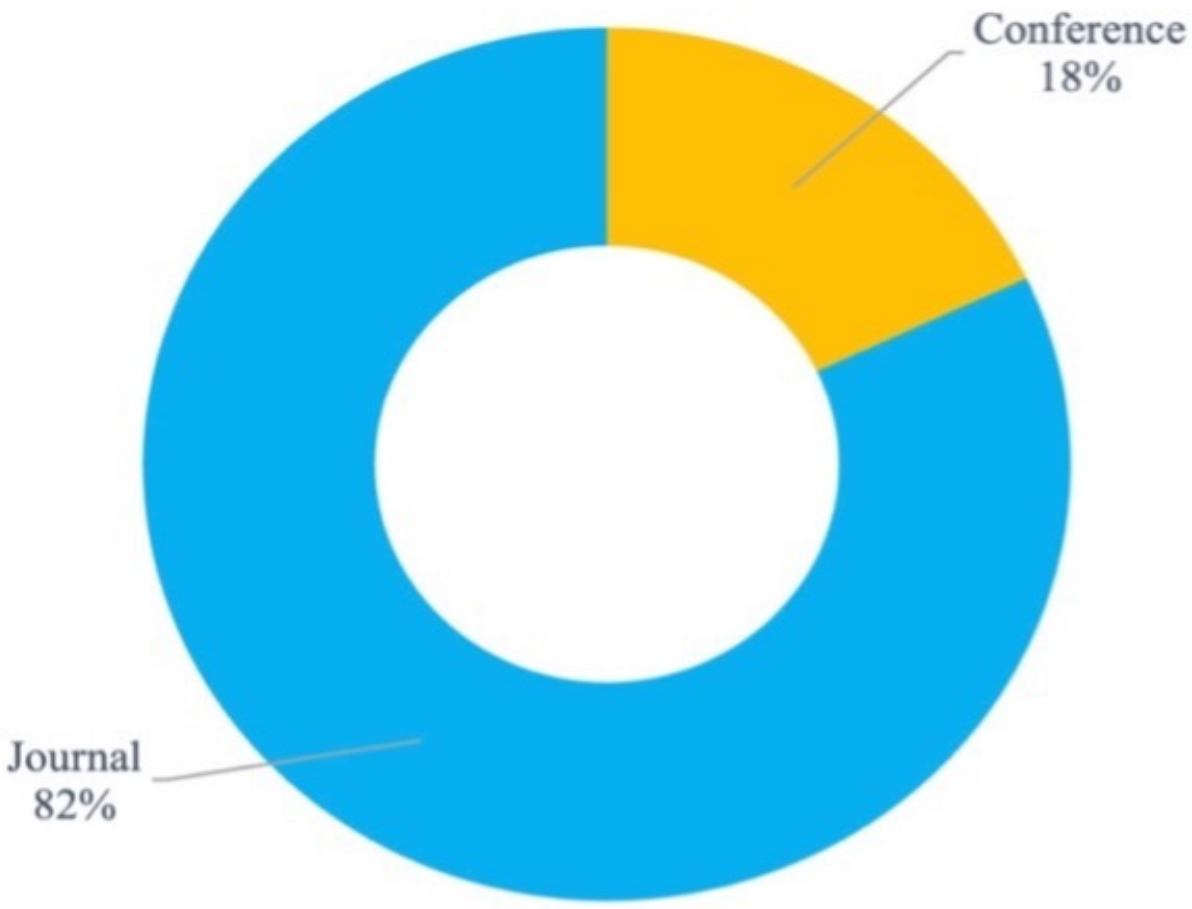

**Fig. 4.** Article types

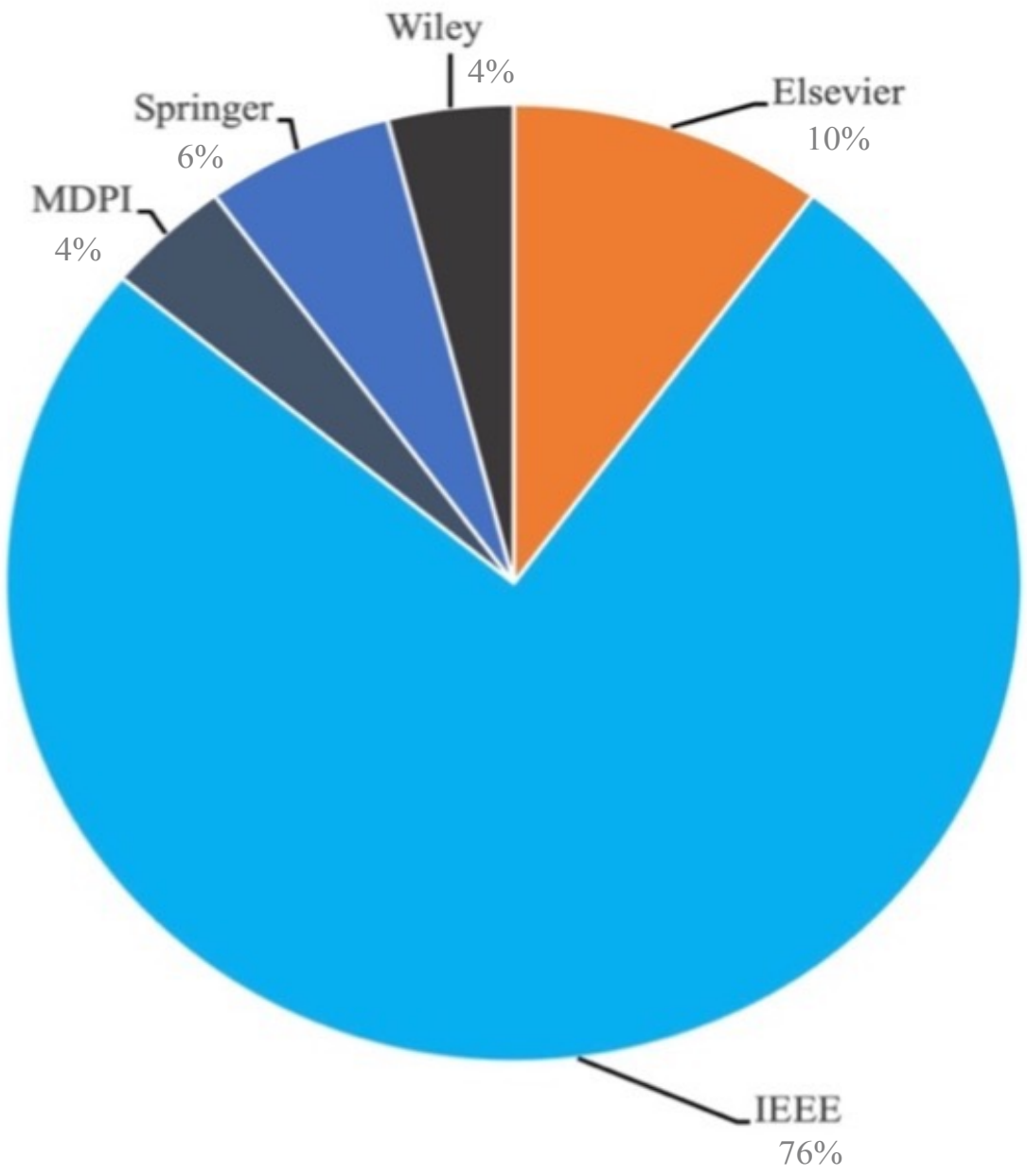

**Fig. 5.** Applied publishers

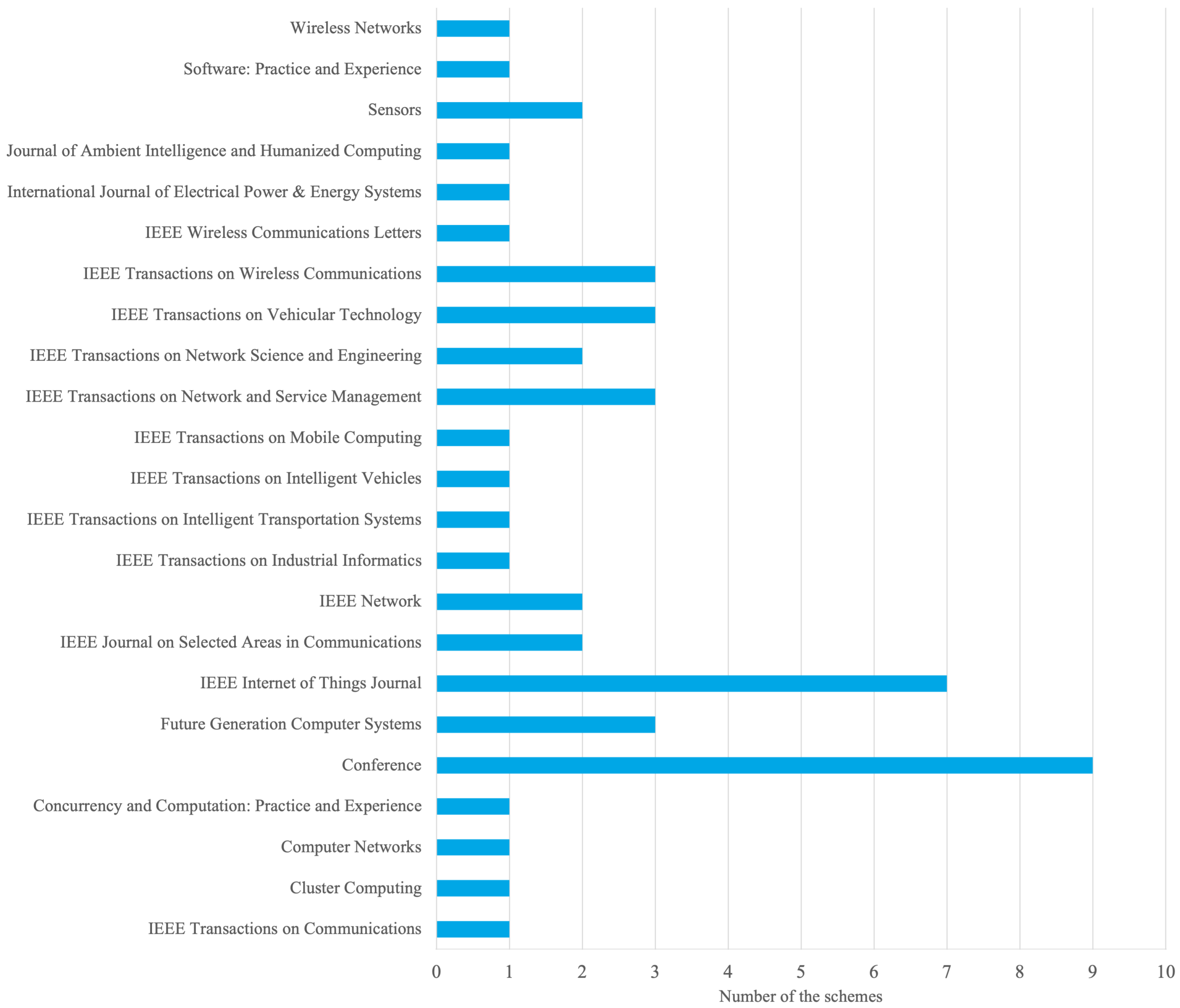

**Fig. 6**. Applied journals and conferences

### 2.2. Questions formalization

The aim of this survey is to examine significant methods and features used in the articles corresponding to a given period of time, as well as the main challenges and problems associated with different schemes for offloading techniques. Considering the primary objective of the survey is to cover the entire topic of MEC offloading within the blockchain context and present related open issues, there are some crucial research questions that need to be addressed.

- How the studied schemes are modeled?
- When it comes to proposed schemes, what type of classification is used?
- What are the most common performance metrics used to evaluate different schemes?
- What simulators are used to assess each scheme's effectiveness?
- What offloading types are considered when assessing the proposed schemes?
- What are the future perspectives and directions regarding the introduced approaches?

In sections 5 and 6, all the above questions have been answered.

## 3. Blockchain

Blockchain technology can keep records and information so they cannot be tampered with, hacked, or changed. Using distributed networks, blockchain facilitates decentralized data storage by chaining blocks in serial and providing a tamper-resistant ledger.

Moreover, by using cryptography, transactions can be recorded and secured [33]. Satoshi Nakamoto proposed the first blockchain in 2008 [23]. which was deployed as the technology that enabled Bitcoin's growth [34]. In addition to digital currency, blockchain has been able to enter most areas to provide security as depicted in Fig 7. As a distributed and secure method for storing data, blockchain technology has proven to be effective. Transactions form the basis of blockchain records. Blockchain broadcasts each new transaction into the entire network as soon as it is generated [35]. By validating the signature assigned to the transaction, the nodes receiving it can check the legitimacy of the transaction, and they can mine verified transactions into cryptographically secure blocks as a result of that validation. A node that creates such blocks is known as a block miner (the short form is miner) [36]. It is necessary for a consensus problem to be resolved in a distributed way in order to enable a miner to create a block. A new block will be broadcast throughout the network by the miners that succeed in solving the consensus problem [37, 38]. As soon as the miners receive a new block and are still unable to solve the consensus problem, they will add the new block to their own chain of blocks which are locally maintained by the miners. Once all the transactions enclosing the block have been verified and the block has also been proven to provide an answer to the consensus problem, they then append the block to their own chains [39]. It is through the use of cryptographic means that the new block is linked back to the previous block in the chain. In order to maintain the consistency of the ledger that's shared across the distributed network, all miners can set up regular times for synchronizing their chains, and specific terms are defined in order to maintain the correct ledger across the network. For example, the Bitcoin blockchain only keeps the chain with the longest chain if there is an error among the chains in the blockchain [40-43] . Functional information about blockchain is shown in Fig. 8.

### 3.1. Types of blockchains

In general, there are multiple forms of blockchains that vary in terms of the data they manage, the accessibility of these data, and the actions they allow the user to take, the following are some of them:

- Public permissionless,
- Consortium (public permissioned),
- Private.

All data in a public permissionless (or simply public) blockchain is visible and accessible to anyone. Nevertheless, certain parts of the blockchain can be encrypted in order to maintain the anonymity of participants [44]. Using a permissionless blockchain, anybody can join the network without any verification and participate in it as a simple miner or node. Economic incentives are usually given to blockchains like these, such as cryptocurrencies such as Bitcoin, Ethereum, and Litecoin [45, 46].

Blockchains of the consortium type enable a small group of nodes to take part in distributed consensus [44]. It can be employed in one or more industries. A consortium blockchain is open for limited public access, and in some cases, centralized within an industry (e.g., the financial sector). Additionally, a consortium between different industries (e.g., financial institutions, insurance companies, and governmental institutions) is still accessible to the general public while still maintaining a partially centralized trust system.

Private blockchains allow only certain nodes to connect. Therefore, it is a centralized yet distributed network [44]. A private blockchain is a permissioned network, which restricts which nodes can execute transactions, run smart contracts, or operate as miners. In addition to being managed by a trusted party, they are also used for private activities. Ripple and Hyperledger Fabric are examples of blockchain frameworks that allow only private blockchain networks [47].

It is also possible to distinguish blockchains based on their purposes:

- In order to track digital assets (for example, Bitcoin).
- In order to run particular logics (for example, smart contracts).

In some blockchains, tokens are used (e.g., Bitcoin, Ripple, and Ethereum), while in others, they are not (e.g., Hyperledger Fabric).

### 3.2. Distributed consensus protocols

A blockchain network needs its peers to be in agreement on a particular state of the distributed ledger as well as the way data should be packed into blocks for it to remain functional. This agreement is known as a “distributed consensus protocol,” which validates the chronological order in which transactions are generated [44]. This ensures that the order in which newly created blocks are added to the shared ledger is agreed upon by a quorum of peers on the blockchain network [48].

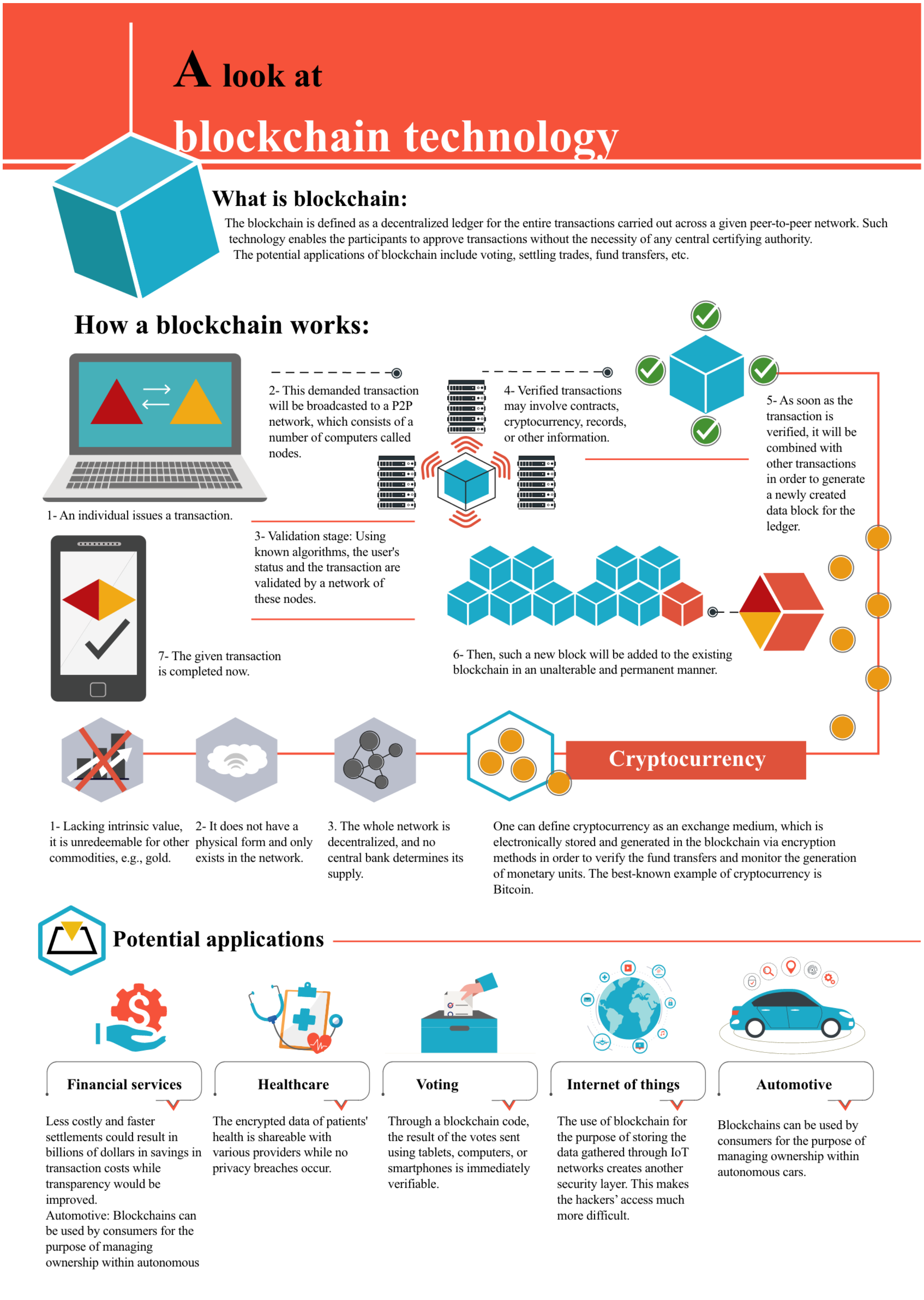

**Fig. 7.** Blockchain steps

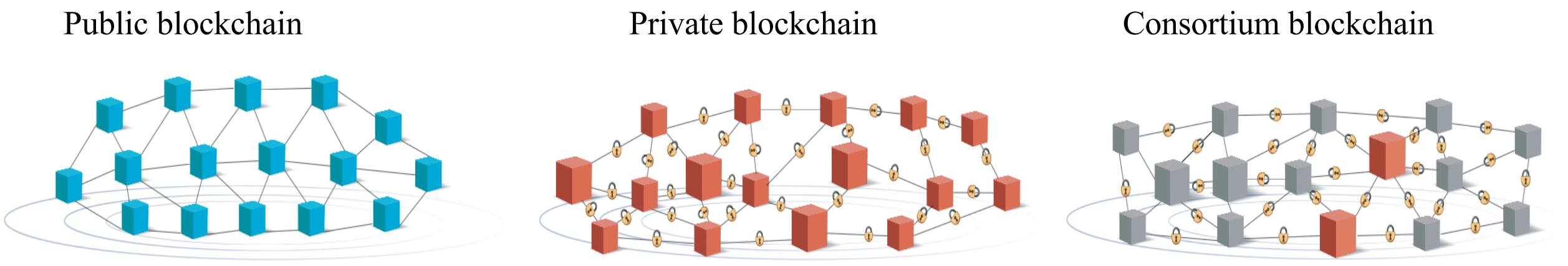

**Fig. 8.** Blockchain types

The following are some examples of distributed consensus protocols [44, 48] :

- Proof-of-work (PoW),
- Proof-of-stake (PoS),
- Delegated-proof-of-stake (DPoS),
- Proof-of-importance (PoI),
- Proof-of-burn (PoB),
- Proof-of-deposit (PoD).
- Proof-of-activity (PoA),

The distributed consensus protocol specifies how a network identifies which peer will process and seal the newly generated block that has yet to be confirmed and formatted. Using a random number generator is the simplest method, but it is ineffective when it comes to network longevity, and in some cases, can even lead to serious consequences if peers decide to attack the entire network at the same time [44]. It is important to know that the idea behind PoS, PoW, and other cryptocurrencies was that each node (miner) contributed something of value to the network, and therefore the best node was rewarded. In addition to promoting competition, the reward also encourages competition between adversaries that enableing them to check each other's work and valuables as a means of preventing a possible attack.

Among the protocols used in the Bitcoin network is the PoW consensus protocol. To determine where the selected peer will be located, computing power is used as a mechanism. Hashing unconfirmed transactions is the method by which peers compete with each other. Consequently, the chance of a peer being chosen is proportional to the computational power of the peer. Each time a peer is chosen and wins, a reward will be given to it. Currently, the Bitcoin network is offering a reward of 12.5 newly generated bitcoins to each peer, which is added to that peer's account when the reward is earned [46, 49] . This process of hashing referred to as mining, is based on the calculation of a block, which contains an unconfirmed transaction link, a random nonce, and a hash reference to the previous block, all of which are combined into one block. The output of the hash must equal a predetermined value in order to be considered valid. A miner, when it achieves the target value, immediately transmits the latest generated block to the network as soon as it has reached that value. In the next step of the process, other peers test the data and if it is correct, it is replicated to the rest of the network [49].

The PoS consensus protocol is designed to make use of the assets a peer has under its control (i.e., the stake in the network value that is controlled by a peer). If a peer is considered a suitable candidate for confirmation of a new block, it is proportional to the number of the peer's assets (i.e., stake or wealth). This is achieved in practice by having a peer deposit a predetermined minimum amount of its assets. As a result, the node is given a ticket to access the network. There is a deterministic pseudo-random selection process where the winner is chosen from a group of peers that have tickets. There is no competition in terms of the computational power of the peers in this case, and that means there is a minimum amount of energy consumption compared to a PoW. However, this approach is similar to that of a shareholder corporation, in which the rich are given the upper hand [47] What makes this approach work is that peer attacks are unlikely to occur since in this case, a peer is attacking its own assets. The PoS consensus protocol is available in several different versions, each of which implements a different method for choosing the validator so as to ensure fairness and consistency in the process, such as Distributed PoS (DPoS). There is a difference between a regular PoS system and a DPoS system, which can be comparable to the difference between a representative democracy and a direct democracy, as stakeholders vote in order for the signers, i.e., the delegates, to be selected [44]. Fig.9, illustrates the difference between PoS and PoW. As a result, it is possible to say that nodes are responsible for distributed consensus, by determining the rules that need to be adhered to. In addition, the blockchain will serve as a mechanism for ensuring that the nodes remain in sync based on the consensus system [47].

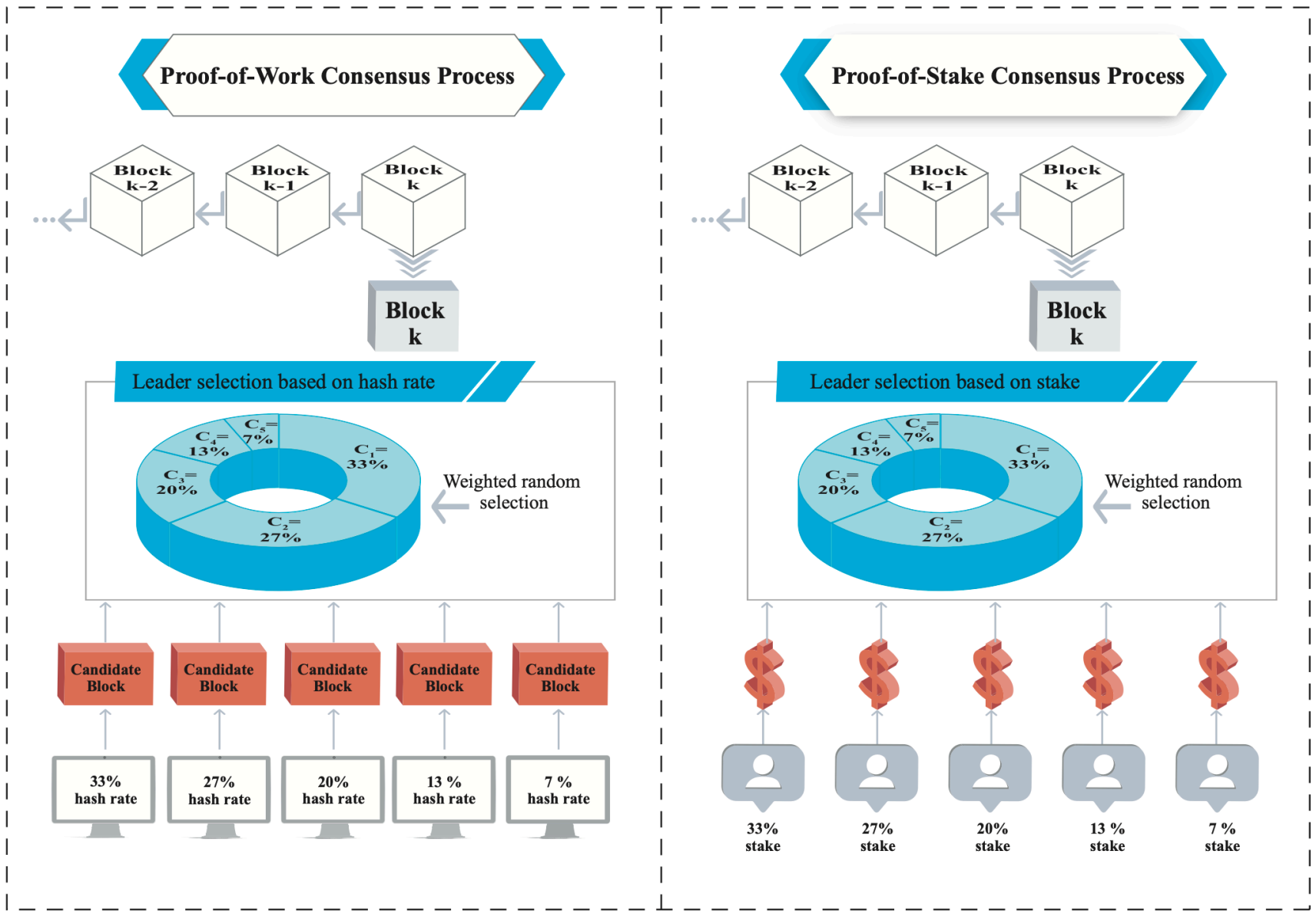

**Fig. 9.** PoW vs PoS

## 3.3. Blockchain characteristics

There are a number of key characteristics of blockchain derived from literature, which is summarized in this section. Moreover, the paper makes a point of highlighting the fact that certain characteristics are defined differently by means of a variety of terminology, which is quite natural for a technology that is still at an early stage. Several characteristics are discussed in the section with the aim of unifying the terminology used for them. It is also discussed what each of these characteristics has to offer in terms of its benefits as well as any problems that may arise. Furthermore, there is a discussion on current issues with Blockchain technology in addition to key challenges for research in this area. It has been identified that blockchain technology possesses the following key characteristics:

### 3.3.1. Decentralization

The term "decentralization" implies the transition of control and decision-making from a centralized entity (an individual, group, or organization) to a distributed network in the blockchain. A decentralized network seeks to lower the threshold for trust and to prevent its participants from exerting control or authority over one another in a way that undermines its functionality [50].

### 3.3.2. Security

Using asymmetrical cryptography, blockchain systems provide inherently secure encryption since public keys are visible to everyone and private keys are only visible to their owners. The keys ensure that the transaction is owned by the rightful owner and that the transaction cannot be tampered with. Keeping transactions confidential, securing their integrity, and authorizing them are the three components of blockchain security [51]. Blockchain systems, which use peer-to-peer consensus mechanisms, eliminate single points of failure for data versus systems that are centrally stored, which are far more susceptible to being compromised.

### 3.3.3. Immutability

In the blockchain, immutability is also known as tamper ability [51], persistency, and unforgeability [44]. In other words, once data has been added to a blockchain, it cannot be changed or altered. Blocks of data in a blockchain structure are timestamped and encrypted using hash algorithms, meaning entry of data into the system is permanent and tamper-proof unless there is consensus amongst the majority of nodes [44]. While the transactions can be seen by anyone at any time, they will not be able to be changed or deleted once they have been validated and added to the blockchain [52]. Changes of any size will generate a new hash and will be detected immediately, keeping the shared ledger immutable [53].Providing or receiving data with this feature ensures that the data has not been altered, resulting in great benefits for financial transactions and financial audits. Nevertheless, immutability has its own challenges and concerns, and some are now questioning its benefits [49].

### 3.3.4. Anonymity

The anonymity feature of blockchain ensures privacy, which is defined as a system of protecting data from unauthorized intrusion or monitoring. In order to achieve anonymity, transactions are authenticated without exposing any personal information about the participants. Through an algorithm establishing trust between nodes, data is exchanged without revealing or verifying information between nodes, allowing the exchange of data to occur anonymously. With a blockchain system, users can hide their real identities by interacting with generated blockchain addresses [49]. Despite its inherent nature as a distributed and public network, the blockchain cannot guarantee complete privacy. As a result, some researchers use the term "pseudonymity" to describe this characteristic of blockchains [49].Blockchain addresses can be created by anyone, and those addresses cannot be connected to people without the information provided by other resources [50].

#### 3.3.5. Democratized

The P2P approach used in blockchain systems allows for democratic decision-making by all nodes [54]. Every decentralized node uses consensus algorithms for the addition of new blocks to a blockchain, as well as for ensuring their copies are synchronized across the nodes and that the block is appropriately appended to the shared ledger [54, 55] , . As a result of the independent nature of all nodes involved in this decision-making process, they possess equal rights and obligations, share data, and assist in maintaining blockchain information, resulting in low maintenance costs over a long period of time [56]. Voting takes place according to the computing power of the node, with valid blocks accepted by trying to extend them and invalid blocks rejected by refusing to be processed [49].

#### 3.3.6. Integrity

The design of blockchain systems makes them inherently resistant to data manipulation. Blockchain data integrity means ensuring the accuracy and consistency of data throughout its entire lifecycle [54]. Blockchain networks achieve this due to their decentralized and virtually immutable shared ledgers, which means that once the transaction record of a block of data is agreed upon and added, its record cannot be modified or edited. Multiplication of copies of this data across the Blockchain network ensures its reliability and integrity, as it is permanently preserved in the Blockchain system [55].

### 3.4. Block structure and content

The Merkel tree binary hash structure is used by the public blockchain networks Bitcoin and Ethereum [57]. In these blockchains, a block is a hash value of information that needs to be logged. The message body and header are the major components of such information [58]. The header usually consists of the block version, parent hash, Merkel root hash, timestamp, nonce, and difficulty level. The block body includes information about transactions packaged as permitted by the particular blockchain [59]. A hash is an alphanumeric value derived from a piece of file or information after it has been processed by a hashing algorithm (for example, SHA-2, SHA-256, and so on). Merkel trees are a widely used approach for protecting data against accidental corruption or intentional manipulation in distributed networks. The technique employs a binary method where a parent node consists of a concatenation of siblings' hash values [60]. Partial verification is the real strength of the Merkel tree technique. The parent hash represents the hash of the predecessor block, and the timestamp represents the time when the transaction occurred [57]. A nonce is a number with a fixed bit length that is solved for values below it by public blockchain miners. The difficulty level is a measure that changes regularly between blockchain networks [57].

## 4. Offloading

There has been a lot of advancement in remote computing over the last few years thanks to the Internet-based paradigms including Edge computing, Fog computing, MEC, and MCC [61, 62] . Moreover, cloud-close paradigms do not meet some of the most important user requirements. As a result of the widespread use of MEC environments to execute modern mobile applications, as well as their role as an essential component of the 5G infrastructure, the MEC architecture has emerged as one of the most attractive architectures within these paradigms to meet the needs of delay-sensitive applications [63]. At the edge of the network, MEC provides computing, data storage, and networking services. It can, in turn, result in a decreased transmission path between the providing and demanding sides, and a reduction in energy consumption and delay for a battery-constrained user equipment. Additionally, it ensures the quality of these services by offloading demanding processes to close-by servers [19, 64, 65].
A typical mobile computing system has the following three major layers:

- The smart devices layer consists of heterogeneous mobile devices in terms of CPU, storage, and capabilities for communication. In addition to communicating with other mobile devices, these mobile devices can also gain access to edge servers (ES) within their coverage range using these technologies. In this layer, the decision is made about which tasks to offload to the remote ESs [28].
- In the edge layer, small data centers with APs and ESs are replaced. A colony of APs scattered throughout the network accesses these servers through optical channels [66]. Connecting the APs themselves is usually done over fiber optics, powerful wires, or wireless channels. Mobile users subscribe to these servers via the nearest APs [67].

- In the fog layer, powerful data centers are located in the APs and ESs. Whenever the above-mentioned layers don't have enough resources to complete the job, the fog layer is used to fulfill the task. Two downward layers can be involved in the decision-making process [28].

A remote MEC infrastructure offloads the resource-intensive tasks that mobile applications must handle to eliminate some limitations [68]. By offloading tasks, mobile users can perform computations faster and reduce the system's total energy consumption and delay [69]. Depending on the type, the request is executed locally or remotely. Local execution is usually more beneficial for lightweight computing applications. A binary offloading or partial offloading decision can be made if remote execution is intended. Once the results have been processed, they have placed into a queue for transmission to the fog or MEC servers in a protected way. Furthermore, some priority algorithms could be set up to queue the requests and send them in response to specific rules.

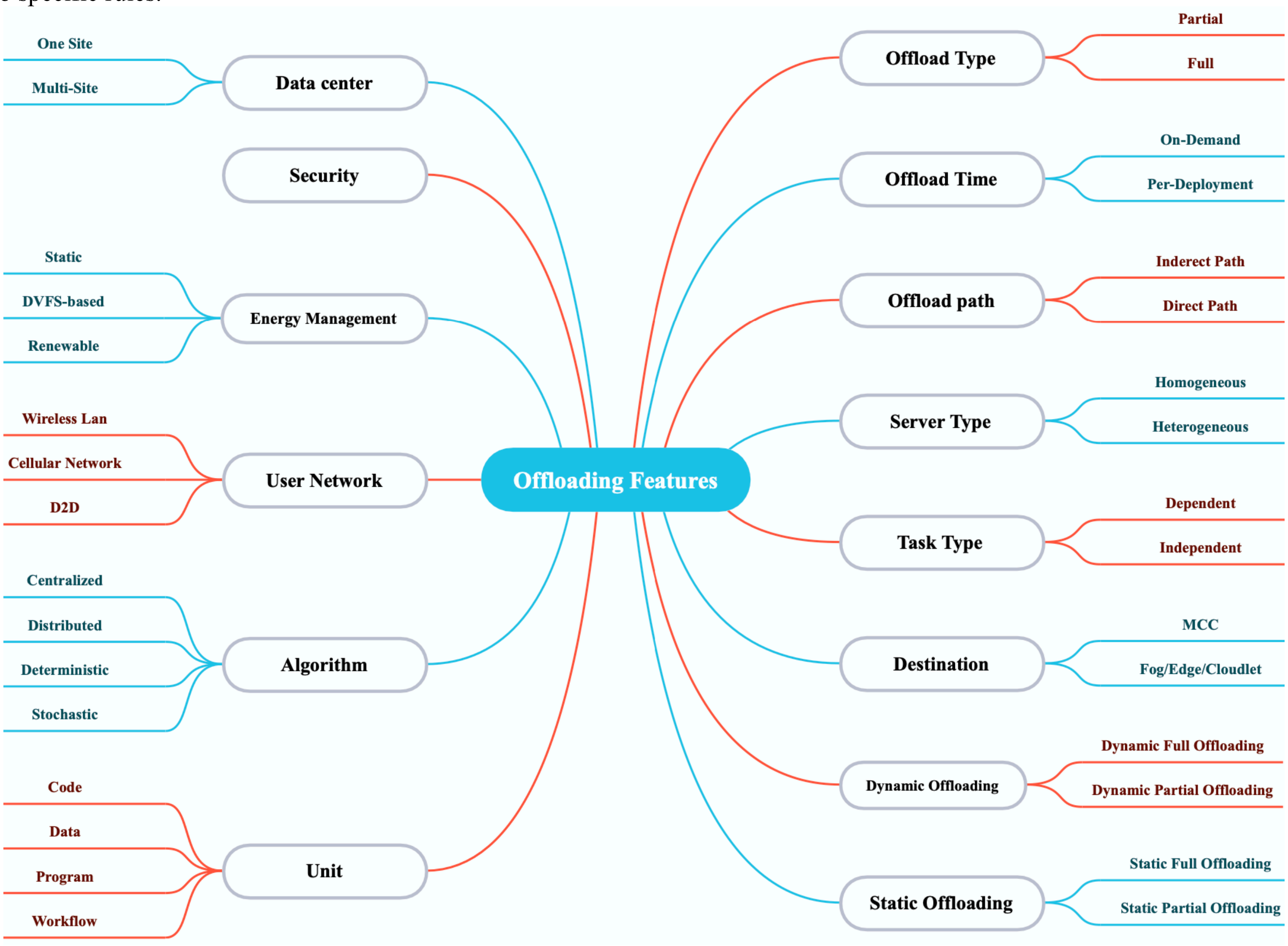

**Fig. 10**. Offloading features

### 4.1. Offloading features

Various characteristics of the offloading process are shown in Fig. 10. For example, offloading schemes can be classified into the following categories based on their granularity [70]:

- Full Offloading: The entire program is offloaded to an offloading server from an SMD. Nevertheless, there might be an increase in network overhead if the program's size is greater than the components' sizes.
- Partial Offloading: The offloading of a workflow or program will only be limited to a subset of it. Interaction overhead between client programs on SMDs and components offloaded on servers should be controlled and may result in runtime overhead.

Furthermore, from a security standpoint, offloading may both decrease and increase security, depending on the application type [71]. In some cases, offloading can increase security if fewer items are transferred. Consider the case of an application on a mobile device that requires data to be stored on the server. In this way, data security will be enhanced, as there will not be any need to send server-side data to the mobile node when the application is offloaded from the server to the mobile node. In contrast, when offloading involves the additional transfer of items that may include data, code, or applications across networks, security can be mitigated .[72]

## 5. Proposed schemes

As Figure 11 shows, one can classify the suggested taxonomy into the following four major fields: Machine learning-based, Game theory-based, NSGA-III-Based and Auction-based.

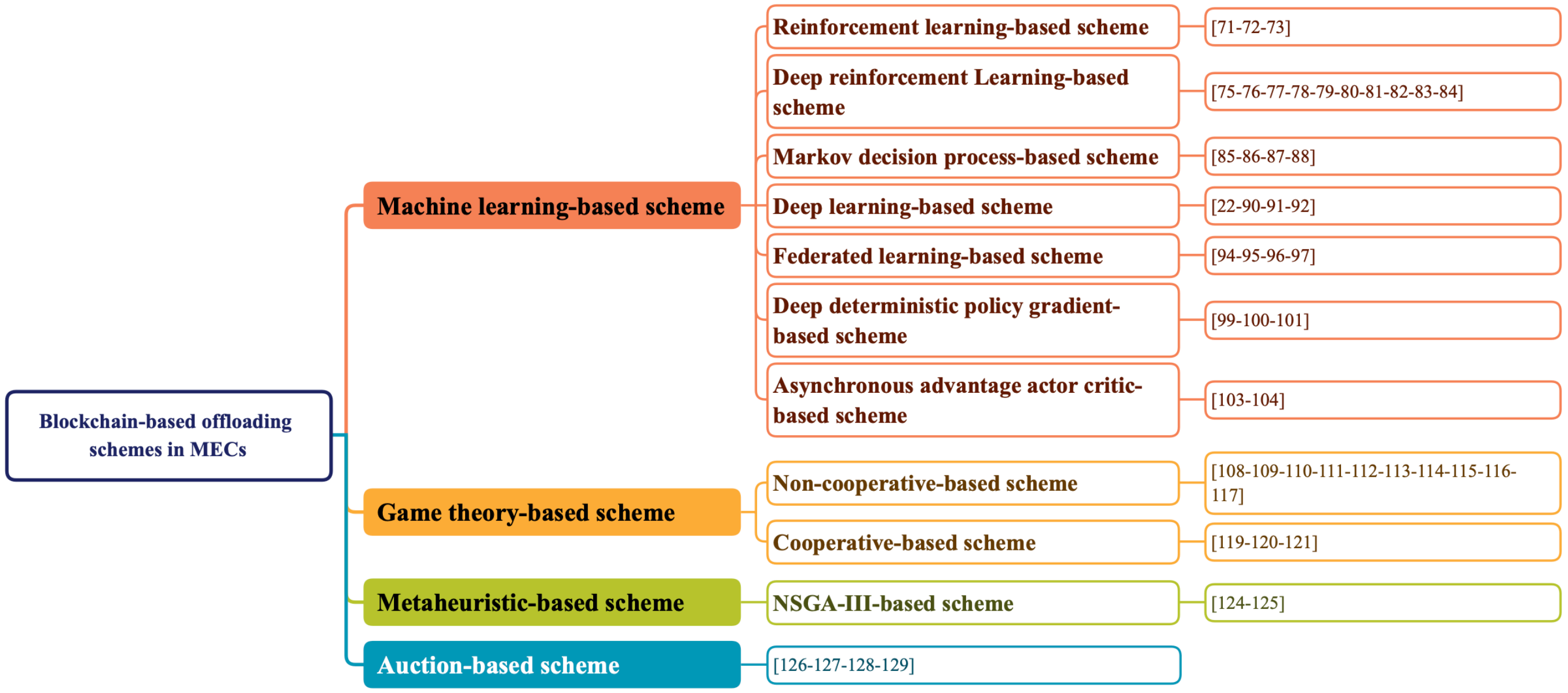

**Fig.11.** Proposed schemes

### 5.1. Machine learning schemes

As a multi-disciplinary model, machine learning is used for extracting knowledge from the input. This is conducted by combining statistics, mathematics, computer science, and artificial intelligence for the purpose of making decisions automatically. Such a decision can be made by learning from the inputs using the following three typical learning techniques in order to reach a specific output without the need for conducting manual manipulations: Unsupervised Learning, Supervised Learning, and RL [29, 73] .

#### 5.1.1. RL-based schemes

By presenting a category of solution techniques to the closed-loop sensory data processing problems so as to create control decisions for reaction, ambient intelligence is introduced into the AIoT systems using the reinforcement learning (RL) technique. In particular, the interactions of the agents with the surrounding environment are carried out for the purpose of learning the optimum policies mapping the status/states to actions [74].

Liu et al. [75], studied the task offloading on the basis of blockchain for the purpose of edge computing on the data characterized by low quality. By considering the limitations of energy consumption and stringent delay, the problem was formulated in such a way that the system utility, e.g., data quality, was maximized. Such a framework is composed of three main constituent elements: evaluation of data quality featuring various data quality dimensions, data repairing by employing repairing algorithms on the basis of a newly developed mechanism of repairing consensus, and distributed RL for task arrangement featuring a distributed RL algorithm on the basis of a newly developed strategy of distributing low-quality data. It should be noted that in the same framework, the quality of shared data can be assessed and enhanced for the purpose of high-quality task offloading policies. By the creation of a balance between data repairing decisions and task offloading decisions, the distributed RL used for task arrangement is capable of providing a flexible method for allocating resources in an edge computing network. The presented numerical results depicted the efficiency and effectiveness of the suggested framework of task offloading for the purpose of edge computing on low-quality data within the Internet of things. In this scheme the authors have provided a comprehensive presentation of the simulations and all evolutionary stages.

Through the modification of the blockchain consensus process, Yao et al. [76], introduced the BC-CED from the currently available solutions. This paved the way for the participants to attain agreements by solving the problem of task offloading. In the first step, they presented the BC-CED designate, which is predominantly composed of the blockchain, resource layers, and an application. Subsequently, they formulated the CED task offloading as a POMDP and showed the procedure of using the RL-based technique for the purpose of solving the same problem. Using the BC-CED, every single participant could employ the RL-based techniques for the purpose of solving such a problem and participating in a competition for the right of block output by making a comparison between the offloading policy performances and the acceptance of the optimum policy as the offloading scheme in the course of the subsequent period. In addition, for the purpose of task offloading, resource management, and functionality of task publishing in BC-CED, these authors have developed four smart contracts and also presented the detailed POO design. In particular, by implementing their POO design in a commercialized blockchain platform BROP, these authors tested its performance in real-world conditions. In order to estimate the BC-CED performance, they carried out some simulation experiments and numerical assessments on the basis of a prototype system. Furthermore, they have presented a truthful incentive mechanism for the purpose of encouraging the contributions of resources in BC-CED and forcing them to behave honestly. In addition, they have presented an incentive mechanism by which trustworthiness is ensured, and also, the CUs are encouraged to contribute their own resources to the BC-CED. The comprehensive experiments conducted by the implementation of the solutions in the context of a commercialized blockchain platform have illustrated how superb performance is achieved by BC-CED in blockchain maintenance and task offloading. In addition, the results obtained at the prototype level have indicated that regarding the load rate, resource usage, and execution delay, their suggested algorithm had a better performance compared to the currently available schemes of offloading. The authors in this scheme have done the simulations and evolution stages comprehensively.

Nguyen et al. [77], introduced a blockchain network on the basis of MEC, in which, by employing wireless channels, multi-mobile users (MUs) could serve as miners offloading their own data mining and processing tasks to the neighboring MEC servers. In particular, a joint optimization problem was formulated that included the preservation of users' privacy, mining profit, and task offloading. Such a problem was modeled as a MDP, which aimed to maximize the blockchain users' privacy levels and minimize the long-term system offloading utility. In the first step, their presented RL-based offloading scheme allowed MUs to make optimum offloading decisions on the basis of the qualities of wireless channel between the MEC server and MUs, the power hash states of users, and blockchain transaction states. Then, to improve the offloading efficiency of larger-scale blockchains further, they introduced a DRL algorithm via a deep Q-network, which was capable of solving large state space efficiently in the absence of prior knowledge about the system dynamics. The offloading performances were evaluated using numerical simulations subject to a variety of conditions for both multiple- and single-user offloading scenarios with regard to computation latency, user privacy, and energy consumption. According to the results of the simulation and experiments, the suggested RL-based offloading schemes resulted in significant improvements in user privacy, reduction of energy consumption, and computation latency while incurring minimum offloading costs when compared to the benchmark offloading schemes. As a limitation, the authors did not generalize the suggested DRL scheme by considering the variations in the action space.

#### 5.1.2. DRL-based schemes

As a promising alternative, Deep Reinforcement Learning (DRL) methods, including Deep Q-Learning (DQL), employ deep neural networks (DNNs) to approximate functions in order to model offloading problems as a MDP [74, 78]. The literature includes a number of suggested RL-based schemes, including [79-83], and in this section, we have reviewed a number of them.

The SDN-based MEC computation offloading service presented by Li et al. [84], aimed to enhance the offloading capabilities and coordination at the control plane. An Entropy-based RL algorithm was presented for the computation offloading of the delay-sensitive tasks at the edge of the networks in order to enhance the exploration degree and deal with changes in a dynamic network. In addition, the process of computational tasks offloading of energy consumption budget and delay toleration has been investigated on the basis of DRL in multi-user MEC. Eventually, their numerical results showed that the suggested E-DRL technique was effective and could outperform baseline algorithms in optimization problems of MEC computation offloading on the basis of the SDN. As a limitation, the authors ignored the contribution of the mobility of user devices. Also, they did not study the offloading problem concerning the joint dynamic MEC computation.

Zhou et al. [85], introduced a DRL-based technique for offloading decision-making and conducted service caching in accordance with both multiple users' offloading decisions and credibility. By the integration of the mechanism of credibility assessment into a smart contract, the credibility was stored within the blockchain for both services and UEs. This could be synchronized among the heterogeneous nodes. They introduced a new technique on the basis of DRL to optimize the overall network efficiency for offloading decision-making and conducted greedy service caching on the basis of the multi-user offloading decisions and credibility. Based on the simulation results, the suggested mechanism exhibited high training efficiency and was capable of reducing the overall delay within the collaborative edge computing network. As a limitation, the authors of this scheme did not explore the potential of using blockchain in wireless networks.

In [86], Samy et al. introduced a framework serving on the basis of blockchain for the purpose of secure task offloading within MEC systems characterized by assured performance with regard to energy consumption and execution delay. In the first step, they introduced blockchain technology as a platform used to attain data integrity, confidentiality, privacy, and authentication of task offloading in MEC. Secondly, by formulating an integration model of task offloading and allocation of resources for a multi-user with multi-task MEC systems, the costs of time and energy were optimized. Given the challenges in the curse-of-dimensionality and dynamic properties of the studied scenario, this could be classified as an NP-hard problem. As a result, a DRL-based algorithm was introduced for efficient close-optimal task offloading decisions. According to the experimental results and theoretical analysis, while the suggested framework remained resilient to multiple task offloading security attacks, it was capable of saving 19.4% and 22.2% of the system's consumption with regard to edge execution and local scenarios, respectively. In addition, based on the benchmark analyses, the framework consumed much lower resources with regard to disk usage and memory, transaction throughput, and CPU utilization. As a limitation, the authors did not generalize their suggested framework so as to jointly consider the pricing of task offloading and a new layer of data compression. Furthermore, they did not consider a more generalized case of a multiuser, multi-server MEC environment.

**Table 4**: Properties of the RL schemes

| Ref | Utilized techniques | Evaluation Parameters | Architecture | Simulators | Key contributions | Advantages |
|---|---|---|---|---|---|---|
| [75] | RL | Data quality, Data repairing in the organizations, The impacts of data quality, The impacts on loss, The impacts on cumulative reward, The performance of DELTA on different devices, The performance of DELTA with multiple actors. | Distributed Task offloading | N/A | A task offloading model for edge computing based on blockchain and distributed RL. | high-quality task offloading policies,<br><br>Flexible method for allocating resources. |
| [76] | RL | CPU occupancy, Memory occupancy, Average bandwidth consumption, Arrival rate, Processing delay,Data throughput. | Distributed Task offloading | N/A | A blockchain-empowered cooperative task offloading for Cloud-Edge-Device computing. | Acceptable system model, Comprehensive simulations and evolution stage. |
| [77] | RL | Processing time, Battery consumption, Total reward, Average latency, Average power, Average offloading cost, privacy level. | Decentralized Task offloading | N/A | MEC based blockchain network for multi-mobile users to act as miners to offload data via wireless. | High blockchain users privacy levels, Reduction of energy consumption. |

In accordance with a strategy of lightweight block verification, Nguyen et al. [87], presented a mechanism of Proof-of-Reputation consensus. They considered a joint design of mining and offloading. By employing a multi-agent deep deterministic policy gradient algorithm, they implemented a new distributed DRL-based technique in order to accommodate the high-dimensional system state space and the highly dynamic environment. According to the experimental results, the superb performance of the suggested TOBM scheme was verified with regard to enhanced offloading utility with much lower latency in blockchain mining, improved system utility, and a better system reward in comparison with the currently available (non-)cooperative schemes. The paper was finally concluded with major technical challenges and the potential directions for the purpose of future investigations on blockchain-based MEC. Among the disadvantages of this scheme, EDs are capable of moving at higher speeds in realistic B-MEC systems in wireless networks (for example, for on-vehicle purposes), which causes extremely dynamic changes in their locations. This affects the offloading decision-making directly. For instance, if an ED is removed from the BS coverage, it will be executed potentially, which causes the offloading design to become ineffective.

Sellami el al. [88], introduced a Blockchain-based DRL technique in order to make it possible to use energy-aware task offloading and scheduling within the Software Defined Networking (SDN)-enabled networks of the Internet of things (IoT). Efficient task offloading and scheduling were attained by the Asynchronous Actor-Critic Agent (A3C) RL-based policy. The latter acted in cooperation with Proof-of-Authority Blockchain consensus in order to validate the blocks and transactions in the Internet of things. In this way, they achieved lower latency, enhanced reliability, and improved energy efficiency for software-defined networking-enabled networks of the Internet of things. The Asynchronous Actor-Critic Agent policy, in combination with the Blockchain, was approved theoretically. They concluded that their technique offered improved scheduling performance and 50% improved energy efficiency compared to the conventional consensus algorithms, such as PBFT and Proof of Work, with regard to network latency and throughput. The investigation conducted by this scheme has not integrated Blockchain transactions with homomorphic encryption.

| **Table 5**: Properties of DRL the schemes | | | | | | |
|---|---|---|---|---|---|---|
| **Ref** | **Utilized techniques** | **Evaluation Parameters** | **Architecture** | **Simulators** | **Key contributions** | **Advantages** |
| [79] | DRL | Average reward, Average delay, Time, Task failure rate,Migration rate. | Distributed Task offloading | Python | A secure blockchain-based mobility handling model for ultra-dense edge computing. | Reduction of delay, Efficient proposed model. |
| [80] | DRL | Latency of blockchain, Throughput. | Decentralized Computation offloading | N/A | A model for wireless services for communication via incorporating DRL and blockchain. | Requires less transmission power, Simplicity of implementing. |
| [81] | DRL | Average system rewards, Total cost. | Decentralized Computation offloading | Python | Trustworthy access control system using blockchain. | Provide high security for MECCO systems, Acceptable system model. |
| [82] | DRL | Average queue length, Average energy consumption, Average Response Time. | Distributed Task offloading | N/A | An energy effective dynamic task offloading model in virtual wireless networks. | Lower computational complexity. |
| [83] | DRL | LOSS, Average cost, Task-drop rate, Average transmit time. | Decentralized Computation offloading | N/A | Model-free DRL-based online computation offloading scheme for blockchain-empowered MEC. | Simplicity of implementing, Strong robustness. |
| [84] | DRL | Energy consumption, Average delay, Normalize reward. | Distributed Computation offloading | N/A | SDN-based MEC computation offloading model to increase the cooperative and offloading power at the control plane. | Effective model in optimization problems of MEC. |
| [85] | DRL | Training loss, Average delay, Offloading ratio. | Distributed Task offloading | N/A | Intelligent edge computing model based on blockchain for cooperative service caching between MEC servers to optimize the resource usage. | Reducing overall delay in cooperative edge computing network. |
| [86] | DRL | Average memory usage, Throughput, Average latency, Max CPU utilization. | Distributed Task offloading | Python | A system for securing task offloading in of MEC based on blockchain using DRL. | High security, Low memory usage. |
| [87] | DRL | Average System Reward, Average Offloading Utility, Block Verification Latency, Average System Utility, Mining utility, Throughput. | Centralized Task offloading | N/A | Collaborative blockchain based task offloading model in order to conducting block mining for improving systems utility. | Simplicity of implementing. |
| [88] | DRL | Latency, Transaction Rates, Block Validation Rate, Throughput, Runtime, Block time, Energy consumption. | Distributed Task offloading | Python | Using blockchain for creating a trustful task offloading model for SDN-enabled IoT networks. | Lowered latency, Increased reliability, Improved energy efficiency. |

### 5.1.3. DRL + MDP based schemes

As a stochastic decision-making procedure, a mathematical framework is employed in the MDP for the purpose of modeling decision-making in dynamic systems. MDP is utilized for random results or those controlled by decision makers making sequential decisions with the passage of time [89]. A number of DRL+MDP-based schemes have been presented in the literature, e.g., [90], and the following subsection has investigated some of them.

To enhance the security and decrease the latency of vehicular CO, Lang et al. [91], presented a blockchain-based handover architecture and mobility-aware CO. In order to optimize the vehicular offloading decisions, they proposed a CO decision problem with models of mobility, blockchain-based handover, and CO. In addition, such optimization was transformed into a MDP, and a multi-agent DRL algorithm was developed in order to solve it. Based on the results of simulation, the benefits of the suggested scheme with regard to latency, reward, convergence, and completion rate of tasks have been approved. As a limitation the authors have not included the vehicle-to-vehicle CO in order to expand the handover and CO scenario.

Zhang et al. [92], presented the elected CBC for the purpose of mediating the caching placement and cache sharing verification. In addition, smart contracts were used to execute the transactions associated with content delivery for the distributed CDM. A consensus protocol was necessary among the smart contract execution nodes in order to avoid fraud and reach a consensus on transactions. In order to learn the associations between user sharing willingness, incentive mechanisms, local traffic offloading, and caching placement, they chose to use DQLN. The performance and reliability of the verification process of transactions were fully realized by the blockchain subsystem. They suggested a partial Practical Byzantine Fault Tolerance protocol for the purpose of minimization of the latency of attaining a consensus, while it guaranteed the confidence level. In addition, the authors formulated the selection of smart contract execution nodes and caching placement as MDP problems. With regard to the dynamics and complexity of the problems, a DRL technique was chosen in order to solve the problems. According to the results of simulation, the enhancement of the suggested scheme was validated. Based on the results, the system was capable of decreasing the latency in content retrieval and offloading the traffic to local caching efficiently by keeping the sharing willingness of the caching nodes at high levels. At the same time, the latency in reaching the transaction execution consensus decreased efficiently in comparison with the traditional Practical Byzantine Fault Tolerance solutions, while the scalability of the system was guaranteed. Although the simulation results of this scheme approved the applicability of the blockchain-incentivized caching system, they did not model the advanced user sharing willingness and allocation of resources on the basis of application.

Also, a multi-agent DRL framework was presented by Li et al. [93], in order to maintain long-term performance for the purpose of cooperative computation offloading. They adopted scattered networks for the purpose of enhancing its stability and introduced league learning for agents in order to collaboratively search the environment for the purpose of maintaining robustness and reaching fast convergence. In the first step, by considering both data processing and blockchain mining tasks, they investigated the offloading problem of non-orthogonal multiple access-enabled cooperative computations. Then, the joint problem was formulated as a MDP. Secondly, in order to prevent unstable performance and fruitless explorations, they primarily employed traditional expert strategies for training an intelligent agent presented as scatter networks. Thirdly, to improve performance, they created a hierarchical league in which the agents cooperated with one another in order to search the environment. Finally, according to their experimental results, their suggested algorithm was capable of presenting improved performance with regard to the reduction of the delay and energy costs and decreasing the training time by up to 60% in comparison with the state-of-the-art techniques. As a limitation, the authors did not study the sparse representation in the field of RL, which could enhance the exploitation as a major problem in DRL.

### 5.1.4. Deep Learning based

As a category of machine learning, Deep Learning (DL) shows much better performance with regard to unstructured data. The popularity of DL approaches, in which deep neural networks are used, is attributable to the increased facilities of high-performance computing. Some of the practical instances of DL techniques include virtual assistants, vision for the purpose of driverless cars, face recognition, money laundering, etc. [94].

On the basis of an algorithm and also the neural network of the MEC scenario, Chu et al. [95], presented a task offloading method and scalable blockchain in order to create a lightweight blockchain so as to prevent the abovementioned drawback for the purpose of minimization of the users' transaction time. Additionally, they presented a technique in order to fast process the consensus in such a way that the suggested technique was capable of achieving higher throughput in comparison with conventional payments via mobile and credit cards. Furthermore, MEC has chosen the verification via offload block for mobile users. On the basis of deep learning (DL), an offloading algorithm has been developed for the purpose of scheduling the block verification tasks for every single mobile user. This led to the minimization of the average execution cost for block verification in the MEC scenario. A number of experimental tests have been carried out to assess the efficiency of the suggested technique. In addition, the efficiency of the suggested technique was analyzed through the estimation of its computational costs. According to the experimental results, the suggested technique showed a high level of scalability and was practically helpful in mobile social networks. As an advantages, the authors have included a complete presentation of the stages concerning the simulations and evolution.

Through secure and effective utilization of aerial vehicles, Masuduzzaman et al. [26], developed an automated real-time scheme for traffic management. In their technique, real-time video data were collected by a UAV from the road junctions. The acquired data were sent to a MEC server in order to be processed. Subsequently, by employing the suggested scheme of two-phase authentication (by combining the cuckoo filter and digital signature), the UAV in BUST was evaluated by the MEC server for a more secure and faster process of verifying the registered devices in the suggested scheme and approving the identity of the UAV. Nonetheless, given the limitations related to the battery capacity and the low computational potential of a UAV, he used multi-access edge computing (MEC) in order to enhance the performance of the traffic management schemes based on an automated UAV. Furthermore, in order to keep the traffic records for the purpose of providing network repudiation and preventing all third-party interferences with the employed network, blockchain technology was included in the automated scheme of traffic management. On the basis of the pairwise compatibility graph concept, an algorithm was constructed for the purpose of the automated UAV-assisted scheme of traffic management in which the vehicles were detected using a DL model. Eventually, the results were analyzed on the basis of the performance and security analyses to decide the effectiveness of the scheme suggested above. As a limitation, the authors did not consider the power charging problem in UAVs.

In [96], Asheralieva et al. investigated pricing and resource management in an IoT system characterized by multi-access edge computing and blockchain-as-a-service (BaaS). Cloud-based servers were included in the BaaS model to carry out the tasks of blockchain. In addition, a collection of peers was included in order to gather data from local devices featuring the Internet of things. The multi-access edge computing model consisted of a collection of aerial and terrestrial base stations (BSs), that is, unmanned aerial vehicles (UAVs), in order to forward the tasks of peers to the blockchain-as-a-service server. Also, every single base station was equipped with a multi-access edge computing server in order to execute a number of blockchain tasks. Given that the base stations were controlled/privately owned by various operators, no data was exchanged between them. They showed that in the BaaS-MEC system, pricing and resource management were modeled as a stochastic game of Stackelberg characterized by several leaders and insufficient data on the peers'/followers' and leaders'/base stations' actions. For the purpose of peers' and base stations' decision-making, they formulated a new hierarchical RL algorithm. In addition, by combining Bayesian DL for peers and DQL for based points, they suggested an unsupervised hierarchical DL algorithm. According to their results, the suggested algorithms converged to stable conditions, in which the peers' actions were regarded as the best responses to the optimum actions of BSs. As an advantages, a complete review of the simulations and evolution stages has been presented by the authors.

Wang et al. [97], introduced a blockchain-empowered and multi-MEC wireless networks of the Internet of things, in which a number of multi-access edge computing servers work cooperatively in order to provide the IoT devices with computation resources so as to realize the consensus mechanisms. In addition, by training a deep neural network via the optimum data, they presented a DL technique. According to the simulation results, the suggested multi-MEC offloading architecture was capable of reaching higher energy efficiency in comparison with the conventional local computing architectures. Additionally, they showed that the DNN-based technique was capable of approaching the optimum results infinitely and also enhancing the efficiency of computations. As an advantage the authors have provided a comprehensive presentation of the simulations and all evolutionary stages.

#### 5.1.5. FL-based schemes

As a distributed collaborative artificial intelligence technique, through the coordination of a variety of devices with a central server FL paves the way for data training without the need to share the actual datasets. Also, it has been utilized in a variety of applications, such as mobile applications, medical applications, transportation, the Internet of things, and Industry 5.0 [98].

In [99], Wang et al. suggested a framework for the purpose of integration of FL, MEC, and blockchain into the IoT system. Nonetheless, the efficiency of the whole system could be hindered by the inadequate throughput of the blockchain. In the meantime, given the dynamic specifications of the MEC and IoT systems, the overall framework involved large-scale actions and high-dimensional properties, and the optimization problem was very challenging and complicated. As a result, by defining the action space, reward function, and state space, the authors designed the overall process of system flow as an MDP sequence. The algorithm of deep deterministic policy gradient was employed so as to choose actions and modify the parameters dynamically so as to solve dynamic and high-dimensional problems. Eventually, according to the simulation results, their scheme enhanced the system's performance significantly. The authors in this scheme have done the simulations and evolution stages comprehensively.

| **Table 6**: Properties of the MDP and DL schemes | | | | | | |
|---|---|---|---|---|---|---|
| **Ref** | **Utilized techniques** | **Evaluation Parameters** | **Architecture** | **Simulators** | **Key contributions** | **Advantages** |
| [90] | DRL | System rewards | Distributed Computation offloading | Python | A model to increase data computation and throughput of blockchain systems. | Low complexity. |
| [91] | DRL | Average reward, Average latency, Average task completion rate. | Distributed Computation offloading | Python | Mobility-aware CO based on blockchain handover architecture to decrease the latency and optimize the security of vehicular CO. | Low complexity. |
| [92] | DRL | Training loss, Traffic offloading, Cache sharing Willingness, Contributed Caching Space, Latency Saving, Consensus Latency, Consensus Reward | Distributed Data offloading | Python | Incentive mechanism for the distributed caching system. | Byzantine Fault Tolerance, Distributed system model, Acceptable traffic offloading and latency. |
| [93] | DRL | Average System Cost, Average Time Cost, Average Energy Cost. | Distributed Computation offloading | Python | Multi-agent deep reinforcement learning model to attain long-term efficiency for collabirative computation offloading. | Efficient algorithm, Reduction of energy cost and training time. |
| [95] | DL | Processing time, Average execution, Transaction per sec. | Distributed Task offloading | NS3 | A model for Task offloading using blockchain technology based on the NL for MEC. | Complete presentation of the stages. |
| [26] | DL | Object detection time, Energy consumption, Authentication time, Throughput, Successful transaction, Average latency. | Distributed Data offloading | Python | Automated real-time traffic management scheme using UAV based on blockchain. | Efficient system model, Reasonable complexity. |
| [96] | DL | Average payoff of the peer, Average payoff of the BS, Average convergence time. | Distributed Task offloading | OPNET | Unsupervised HRL and DL system for a stochastic Stackelberg game with numerous users under incomplete information. | Good presentation of the proposed model. |
| [97] | DNN | Sum of Energy Efficiency, LOSS, Average Energy Efficiency | Distributed Computation offloading | N/A | Blockchain empowered and multi-MEC wireless IoT networks for realizing consensus mechanisms. | Multi-MEC offloading architecture, Simplicity of implementing. |

Also, a normalization method was developed by Zhao et al. [100], The authors showed experimentally that in cases where features were subject to differential privacy protection, their normalization method performed better in comparison with batch normalization. Furthermore, an incentive mechanism was designed to give rewards to participants so as to attract more customers for participation in the crowdsourcing FL task. In the course of the federated training, the blockchain audited the updates pertaining to all customers. In this way, the system was capable of holding the model updates accountable so as to avoid malicious manufacturers or customers. The system presented in this scheme was not adequately tested by employing the datasets obtained from real-world home appliances. In addition, for the purpose of presenting better test accuracy, a deterministically optimal balance needed to be established between the global and local epochs.

In addition, a blockchain-controlled, edge intelligence FL framework was presented by Alghamdi et al. [101], for a distributed learning platform for the purpose of consumer IoT data analyses. The FL platform paved the way for collaborative learning using the data shared between users, and the centralized aggregator was replaced by the blockchain network, which guaranteed the secure contribution of gateway devices within the ecosystem. The overall security challenges in an ecosystem, e.g., data leakage scopes in the gateway, were taken into consideration. In addition, the blockchain was immutable, anonymous, and trustless, which encouraged the participation of the CIoT end users. Also, they used a consensus procedure for the creation of a global model in order to extend the security of model sharing and accelerate the learning process. In accordance with the findings of the extensive evaluations obtained from real-world datasets, the best method of model selection on the basis of the consensus led to safety improvement and showed significant output differences in comparison with the conventional FL techniques. The authors used the well-known Stanford Cars dataset to analyze the framework and FL outcomes. The effectiveness of the suggested framework was approved by the experimental results. As a limitation, this scheme is lacked a blockchain protocol-independent framework for the purpose of federated machine learning.

Also, an MCS FL system was presented by Wang et al. [102], on the basis of edge computing and Blockchain for the purpose of resolving problems of privacy protection in MCS. In particular, this paper employed the FL as the MCS system framework. The mentioned system used Double Local Disturbance-Localized Differential Privacy presented in the same article to protect the location and data privacies. Given the fact that the sensed data was available in a variety of modalities (e.g., video, audio, text), the paper employed the Multi-modal Transformer technique in order to combine the multi-modal data prior to conducting the subsequent operations. In such a system, the aggregation tasks and model training were offloaded to the edge server in order to resolve the inadequate computing problems on the mobile terminal and improve the efficiency of data processing for the platform of crowdsourcing. In order to resolve the problem related to the untrusted label for third parties, the blockchain was used for the purpose of distributing the tasks and collecting the models. They presented a reputation calculation (Sig-RCU) technique in order to determine the real-time reputation of the participants of the designated task. The adaptation and effectiveness of the suggested Double Local Disturbance-Localized Differential Privacy and Sig-RCU algorithms were approved by the implementation of experimental tests on the real datasets. Designing an effective incentive mechanism in this scheme was necessary in order to enhance the service quality and enthusiasm of the task participants. In addition, there was a necessity to integrate other state-of-the-art technologies into the mechanism of privacy protection in MCS in order to decrease data quality loss and enhance privacy protection potentials.

#### 5.1.6. DDPG-based schemes

The Deep Deterministic Policy Gradient (DDPG) algorithm is regarded as a model-free off-policy actor-critic algorithm in which the DQN and DPG ideas are combined. Lillicrap et al. (2015) was the first author to present the DDPG algorithm [103].

In [104], Hu et al. developed a DDPG-based algorithm for the purpose of solving the problem of the MDP and defining the variable and multiple numbers of successive time slots as a decision epoch in order to train the model. In particular, DDPG was capable of solving a problem of the MDP with an uninterrupted action space while requiring only a straightforward actor-critic architecture, which made it favorable for the purpose of managing the complexity and dynamics of the MEC-enabled blockchain systems. According to the simulation results, the suggested scheme showed better performance in comparison with DQN-based schemes and a number of other greedy (non-joint optimization) schemes subject to various LTF thresholds, weight factors, and task arrival rates. The joint optimization scheme was capable of reaching improved performance in comparison with other schemes characterized by a higher rate of convergence. In addition, the authors debated the contribution of the number of mobile users to the convergence efficiency, and the joint optimization scheme constantly showed some benefits over the alternative schemes. In sum, their suggested scheme made it possible to deploy blockchain technology in networks of the Internet of things efficiently, and as a result, it could be used for security-sensitive and latency-sensitive IoT applications. As an advantage the authors of this scheme have included a complete presentation of the stages concerning the simulations and evolution.

TOBM, which was developed by Nguyen et al. [105], is a cooperative block mining and task offloading scheme for blockchain-based multi-access edge computing systems, in which besides managing data tasks, each edge device also manages block mining in order to improve the system's utility. In order to overcome the latency problems resulting from the blockchain operation in multi-access edge computing, the authors developed a novel mechanism of Proof-of-Reputation consensus on the basis of a lightweight block verification strategy. This was followed by formulating a multi-objective function for the purpose of the maximization of the system utility in the blockchain-based multi-access edge computing systems through joint optimization of the channel selection, offloading decisions, and allocation of computational resource transmit power. By employing a multi-agent DDPG algorithm, the authors presented a distributed DRL-based technique. Then, a game theoretic solution was presented in order to model the mining and offloading competition between the participant edge devices as a potential game, which approved the presence of a pure Nash equilibrium. The suggested technique could be potentially used for future intelligent mobile networks, in which EDs would be capable of building distributed intelligent solutions through their cooperative DRL model so as to enable intelligent communications, network control, and computation. As a limitation, the authors did not study the tradeoff between latency and mining security with regard to blockchain, which could establish an advantageous balance between the above key design parameters before their integration into MEC.

**Table 7**. Properties of Federated learning schemes

| Ref | Utilized techniques | Evaluation Parameters | Architecture | Simulators | Key contributions | Advantages |
|---|---|---|---|---|---|---|
| [99] | FL | Reward | Decentralized Computation offloading | N/A | Reliable task offloading optimization model based on DDPG to ensure security in IoT devices. | Optimal system performance, Low complexity. |
| [100] | FL | Test accuracy, Reward value, Training time, Reputation value. | Centralized Data offloading | N/A | A model for customers better learning based on blockchain and FL system for IoT devices manufacturers. | Differential privacy protection methods, providing high level privacy for protecting customers' data. |
| [101] | FL | Loss, Accuracy, Time. | Distributed Task offloading | Python | federated machine learning system based on blockchain for consumer IoT data analysis. | Low complexity. |
| [102] | FL | Data quality loss, Algorithm running time, Accuracy, Upload data size, Merkle tree generation time, predicted similarity, Average real service quality. | Distributed Task offloading | Python | MCS system based on FL to protects location and data privacy. | Efficient model to solving privacy leakage problem. |

The intelligent computing offloading model of the IoV on the basis of blockchain presented by Qi et al [106], not only verifies the IoV security but also realizes the low-cost, efficient, and reliable offloading of intelligent vehicular computing tasks. In order to minimize the overall cost of computing offloading, the paper examined the computing roles of intelligent vehicles and blockchain thoroughly, established a computing offloading model subject to different limitations, including energy consumption and time delay, and employed the DDPG algorithm in order to realize the computing offloading scheme so as to minimize the system utilization costs. According to the experimental results, the intelligent computing offloading model of the IoV developed in this work was capable of reducing the overall cost of the model effectively, while the computing offloading success rate was improved. As a limitation, the authors did not consider the optimization of blockchain throughput adequately in order to optimize the IoV computing offloading subject to more limitations.

#### 5.1.7. A3C-based schemes

The A3C algorithm utilizes a parallelized asynchronous training scheme (via multiple CPU threads) to improving efficiency. A3C is regarded as an on-policy RL technique in which no experience replay buffer is employed. The A3C algorithm allows several workers to make interactions with the environment simultaneously so that the local gradients are computed. The computed local gradients of each worker are passed to a global neural network, which is responsible for conducting the optimization and synchronizing with the workers asynchronously [107].

In addition, a cooperative resource allocation and computation offloading framework was presented by Feng et al. [108], for blockchain-enabled multi-access edge computing systems. A multi-objective function was designed in this framework in order to maximize the transaction throughput and the computation rate of the multi-access edge computing systems of blockchain systems through joint optimization of the power allocation, offloading decisions, block interval and block size. Given the dynamic properties of the processing queues and the wireless fading channel at multi-access edge computing servers, one may formulate the joint optimization as a MDP. A resource allocation and cooperation computation offloading algorithm on the basis of the A3C was designed in order to solve the problem of the MDP for the purpose of dealing with the complexity and dynamics of the blockchain-enabled multi-access edge computing systems. The same algorithm optimized DNNs via asynchronous gradient descent and elimination of the data correlation. According to the simulation results, the suggested algorithm could reach convergence faster and attain noticeable performance enhancements over the available schemes with regard to the overall reward. As a limitation, interference management was not studied in blockchain-enabled MEC systems.

Also, an edge-terminal collaborative mining task processing framework was developed by Xu et al. [109], in order to enhance the blockchain system's computing potential. To resolve the problem of insecurity and low efficiency in microgrid transactions, blockchain, and edge computing technology were integrated in order to suggest an edge-terminal collaborative mining task processing framework, which gives users the permission to offload mining tasks locally to D2D users/edge nodes, and the intelligent devices. Such a framework comprised the following three working modes: user collaboration, edge node collaboration, and local computing. In particular, in order to prevent the security threats resulting from malicious nodes, they considered the trust value of collaborative user nodes. In addition, a delay-and-throughput-based blockchain computing task offloading model was established, and an A3C algorithm was utilized in order to jointly optimize the allocation of transmission power, offloading decision, size configuration, and block interval. According to the simulation results, the suggested algorithm was capable of reducing the average delay by 2.5% and 1.7% in comparison with the Fixed-BlockSize and Only-MEC algorithms, while the average transaction throughput was enhanced by 28.5% and 12.1%, respectively. As a limitation, the mobility and also the differentiated processing potential of the terminal, which is favorable for practical purposes, have not been investigated.

**Table 8**: Properties of the DDPG and A3C schemes

| Ref | Utilized techniques | Evaluation Parameters | Architecture | Simulators | Key contributions | Advantages |
|---|---|---|---|---|---|---|
| [104] | DDPG | Average reward, Average power, Average block interval. | Distributed Computation offloading | Python | MEC-enabled blockchain Model for IoT networks in order to optimization blockchain transaction throughput and MEC computational output. | Effective joint optimization scheme, Lowest power consumption. |
| [105] | DDPG | Average system reward, Average System Utility, Block Verification Latency, Average Offloading Utility. | Distributed Task offloading | Python | Collaborative TOBM model to enable a joint scheme of blockchain mining and task offloading in MEC-enabled blockchain systems. | Multiple algorithms and methods, Efficient offloading utility, Lower blockchain costs. |
| [106] | DDPG | LOSS, Reward, Total cost, Success rate. | Computation offloading | Python | Blockchain based intelligent computing offloading scheme for the IoV. | Reduction of the total cost. |
| [108] | DDPG | Average reward, Average computation rate, Average transaction throughput | Distributed Computation offloading | Python | blockchain-enabled MEC model. | Multi-objective function framework, Efficient algorithm. |
| [109] | A3C | Loss connection, Reward, Time delay, Average transaction throughput. | Centralized Task offloading | Python | Edge-terminal cooperative mining task processing system | Multi-objective joint optimization, Acceptable delay, Better transaction throughput. |

## 5.2. Game theory-based schemes

The game theory dates back to the early 20th century when the min-max theorem was approved. It has been utilized as a robust technique for the analysis and solving of problems in the field of social science economics [110], etc., featuring natural competition. A few efforts have been made in the field of computer science in order to apply this theory to solve the problems in resource allocation and job scheduling [111]. In general, it is required to divide a problem into three main characteristics, i.e., the players, the strategies adopted by them in a formulated game, and the utilities of every single player.

#### 5.2.1. Non-cooperative-based schemes

In general, noncooperative games are aimed at achieving Nash/Stackelberg equilibrium as their solution. The noncooperative game includes some players with a number of interests shared among them as a result of the decision-making process results. The same category of the game theory is describable via the final decision units composed of the individual players, unavailability of commitments, and complete rules [112]. Given that the noncooperative game theory is completely matured, one may categorize it into various subdivisions, a number of which have not been scrutinized in the investigated papers. In general, the most widely used Game theories include stochastic, Stackelberg, and potential game theories, which have received the attention of many researchers. A number of non-cooperative-based schemes, including [99, 102, 113-115] have been presented in the literature, some of which have been investigated in the present section.

In [116], Zhang et al. introduced the banks, which were capable of providing loan services to the MEs in order to address the abovementioned two problems. In their work, the problem was formulated as a non-cooperative game in order to model the competition among the myopic MEs. They used a potential game technique to prove the presence of pure-strategy Nash equilibrium (NE) and developed a distributed algorithm in order to attain a Nash equilibrium point with much lower computational complexity. In addition, using theoretical proof, they presented an upper bound on the price of the game anarchy. Furthermore, two smart contracts were presented in order to carry out the coin loaning and computing resource trading processes automatically. In addition, the execution of the two smart contracts required only low financial costs on the Ethereum network. Also, they designed a smart trading contract for automatic trading of the computing resources between edge servers and MEs and a smart loan contract for the purpose of automatic repayment and loan between banks and MEs. Finally, according to their simulation results, the presented algorithm was capable of reducing the overall cost of the whole MEs significantly, and it outperformed other alternative solutions and scaled favorably upon the increase in the number of MEs. As a limitation, the interests of the edge servers, banks, and MEs were not optimized simultaneously using the Stackelberg game in this scheme.

Zuo et al. [117], introduced a multi-access edge computing server to introduce mobile blockchain networks in which the whole mobile users participated in the proof-of-work (PoW) mining activity. In order to keep a stable block time of mobile blockchain networks, the authors formulated a number of delay-limited computation offloading strategies for the purpose of proof-of-work-based mining tasks as a non-cooperative game, which maximized the individual revenues in the mobile blockchain networks assisted by MEC. As the next step, they particularly analyzed the problem of sub-game optimization and approved the NE existence for the same non-cooperative game. In addition, they developed an alternating iterative algorithm on the basis of greedy rounding and continuous relaxation in order to reach the Nash equilibrium of the game. With regard to the optimal computation offloading strategies, the optimum transmission power was also determined for individual users within the maximum range of mining delay. The suggested algorithm was capable of attaining the optimum delay-limited computation offloading and transmitting the power strategies for every single user efficiently. For all users, the individual transmission power has extended in accordance with the optimum strategies of computation resource allocation. There are a variety of parameters, including block size, CPU frequency of the multi-access edge computing server, and the number of users, which can affect the system performance of the suggested delay-limited mobile blockchain networks greatly. The authors in this scheme could employ the cluster cooperation approach in order to investigate the contribution of multiple MEC servers-assisted mobile blockchain networks to block storage and computation offloading.

In the first stage of their investigation, Zuo et al. [118], presented an untrusted multi-access edge computing PoW scheme in mobile blockchain networks that could offload a great number of nonce hash computing demands to the multi-access edge computing server. Afterward, they designed a nonce ordering algorithm for the same scheme in order to present a fairer mechanism of computing resource allocation for the whole mobile IoT users or devices. In particular, the nonce selection strategy of the user was formulated as a non-cooperative game, in which, for every single user, the utilities were maximized in the untrusted networks of MEC-aided mobile blockchain, and the Nash equilibrium was regarded as the problem solution. In addition, the NE existence was analyzed, and by employing repeated games, it was shown that the cooperation behavior was unfavorable for the IoT devices enabled by blockchain. In addition, the authors showed that it was not feasible to use the cooperation technique of the repeated game for the users of the Internet of things in the mobile blockchain networks. Eventually, the mechanism of difficulty adjustment in the blockchain was developed in order to become assured of the stability of block times during long time periods. The suggested nonce ordering algorithm was capable of providing optimum nonce selection strategies and fairer computation resources for every single mobile user when compared to the weighted round-robin algorithm. Using the suggested mechanism of the difficulty adjustment in the blockchain, the network became stable. As an advantage in this scheme, a complete review of the simulations and evolution stages has been presented by the authors.

By employing a prospect theoretical method, Zhang et al. [119], designed an optimal offloading scheme used for MEC-empowered blockchain that considered the varied risk and profit preferences of devices explicitly. In particular, the offloading process was formulated as a Stackelberg game incorporating the notions adopted from the prospect theory. In addition, they approved the NE existence for the games and presented an efficient offloading algorithm for the mining task. They developed an efficient algorithm for the purpose of reaching the optimum strategies for offloading that led to the maximization of the utilities for both the multi-access edge computing service providers and the miner devices. The performance of the suggested offloading schemes was validated by the presentation of numerical results. As an advantage, the authors of this scheme, have provided a comprehensive presentation of the simulations and all evolutionary stages.

In [120], Guo et al. suggested that one can select the free resources shown on edge cloud and non-mining devices in order to create a collaborative mining network (CMN) so as to implement the mining tasks for mobile blockchain. If the resources were inadequate, the miners could offload their mining tasks to non-mining devices in the edge cloud or a collaborative mining network. Given the competition for resources among the non-mining devices, they formulated the problem of allocation of resources in a collaborative mining network as a double auction game, in which Bayes-Nash Equilibrium (BNE) was analyzed in order to determine the optimum auction price. After completing the offload to the edge cloud, they adopted the Stackelberg game in order to model the interactions between various collaborative mining networks and edge cloud operators to reach the optimum price for resources and meet the resource demands of devices. Such a mechanism improved the mining utility in the mining networks while the maximum edge cloud operators' profit was ensured. Eventually, they compared the mining networks' profits with those of an existing mode, in which only the offloading to the edge cloud was considered. Given the suggested mechanism, on average, mining networks could obtain 6.86% more profits. As an advantage, the authors of this scheme have included a complete presentation of the stages concerning the simulations and evolution.

#### 5.2.2. Cooperative-based schemes

Cooperative game theory is concerned with the cooperation between rational users based on a predefined agreement between them. The same type of cooperation influences the choices of users and, thereby, their utilities. One may characterize this category of game theory via implicit and broadly defined regulations, availability of commitments, and emphasized coalitions [112]. In general, cooperative game theory can be categorized into two major subclasses: coalitional games and bargaining theory. In the latter, players with shared resources and conflicting interests are capable of attaining a mutual benefit as a result of the bargaining situation context, where players are not required to approve an agreement [121].

In [122], chen et al. introduced a multi-hop distributed and cooperative algorithm for computation offloading, in which the mining and data processing tasks are considered simultaneously for the purpose of blockchain-empowered IIoT. The offloading problem was formulated as an MCOG, and the presence of NE was approved for the game. Subsequently, a high-efficiency distributed algorithm was designed on the basis of the finite enhancement characteristic of the MCOG. Eventually, according to the experimental results, the suggested BU algorithm could reach a fast-paced convergence to a stable state. Also, its number of iterations increased almost linearly as the number of UEs increased. Such a result indicates that as the number of UEs increases, their suggested algorithm can scale well. Also, the algorithm showed a comparatively high-efficiency subject to a variety of parameter settings. The authors have included a complete presentation of the stages concerning the simulations and evolution.

Using blockchain technology, Lang et al. [123] reached efficient data sharing between the service providers (namely, server vehicles) and vehicles and ensured the computation offloading security among vehicles. In the first step, a consensus mechanism employed in blockchain that combines Practical Byzantine Fault Tolerance and Proof of Service was developed in the same architecture in order to enhance data sharing efficiency and security. In addition, by employing an offloading game, a cooperative offloading decision-making technique was suggested, and by employing the same technique, NE was achieved for the offloading strategy. The advantages of the suggested technique concerning latency were affirmed by the performance assessment. A user vehicle could decide whether to delegate its own tasks to a nearby server vehicle or a roadside multi-access edge computing server via the computation offloading strategy. As a result, the suggested technique could be of help in enhancing the efficiency and security of cooperative computation offloading in vehicular edge computing networks serving on the basis of blockchain. As a limitation, with regard to blockchain, the authors did not investigate the tradeoff between latency and mining security so as to establish an advantageous balance between the same key design parameters before integrating them into MEC.

In [124], Zuo et al. presented a cooperative (multi-access edge computing)-aided blockchain network. In this network, devices were capable of offloading computation-intensive proof-of-work mining tasks to BSs and using the cloud service provider for storing their own block data. In the same network, the interaction process between base stations, CSP, and IoT devices was formulated as a three-stage Stackelberg game. Subsequently, the joint problem of computation offloading, resource service pricing, and block storage was studied as a three-stage Stackelberg game.

**Table 9**: Properties of the noncooperative Game theory schemes

| Ref | Utilized techniques | Evaluation Parameters | Architecture | simulators | Key contributions | Advantages |
|---|---|---|---|---|---|---|
| [99] | Stackelberg | Utility of the requester, Average utility of the providers, Average energy, Time. | Distributed Computation offloading | N/A | Joint model of computation offloading and computing resource trading based on blockchain in D2D-assisted MEC to motivate users to take part in resource sharing. | Good presentation of the proposed model. |
| [113] | Stackelberg | The convergence performance, Number of PMs, The execution time, The $U_t$ vs. iterations. | Distributed Computation offloading | N/A | Cooperative computing tactics by leveraging computing resources on multiple PMs based on an MEPC. | Optimal service pricing. |
| [114] | Stackelberg | Price paid per block, Required transaction rate, Transaction per second, Necessary confirmation blocks, Confirmation delay. | Distributed | N/A | Stackelberg game for blockchain users and the blockchain miners for paying fees, E-Health monitors. | Low complexity. |
| [102] | Stackelberg | Utility, Reputation, Total cost of user, Task delay. | Task offloading | N/A | Task offloading between master node, the users, and validation nodes as a three-stage Stackelberg game. | lower delay, Low complexity. |
| [115] | Stochastic | The total net revenue, The net revenue of computing, The probability of selecting mode, Task offloading decision mode, The net revenue of computing, Average delay. | Distributed Computation offloading | N/A | MEC-enabled wireless blockchain system to handling the problems of PoW puzzle. | Multiple offloading modes considered. |
| [116] | Potential | Number of successful convergences, Number of iterations, Time slot, System cost. | Distributed Computation offloading | N/A | Modeling an effective distributed algorithm based on a potential game theory. | Distributed algorithm, Reduction of the total cost. |
| [117] | Nash equilibrium | Nonce Length, Offloading Ratio, Individual Revenue, Average Revenue, Average run time for MEC servers, Transmit Power. | Distributed Computation offloading | N/A | Single computation resource distribution tactics as a non-cooperative game. | Multiple sub-game optimization, Fast convergence, Good stability. |
| [118] | Nash equilibrium | Average Nonce lengths, Block time. | Distributed Computation offloading | N/A | Nonce ordering algorithm to provide fairer computing resource allocation for all IoT users. | Efficient model, Reasonable complexity. |
| [119] | Stackelberg | Total utility, Service price. | Distributed Task offloading | N/A | Computing resource pricing and utilization as coupled leader cohort games. | Optimal offloading scheme, Simplicity of implementing. |
| [120] | Stackelberg | Mining success possibilities, Average auction price, Optimal price of ECO, Optimal average resource, Comparison of CMNs' profit, Profit of EC, Mining profit of CMN. | Distributed Task offloading | N/A | Free resource displayed on non-mining- devices and edge cloud to execute mining tasks based on mobile blockchain. | Obtain more profits. |

In each stage, the authors analyzed the subgame optimization problem. They presented an iterative algorithm on the basis of the backward induction to reach NE for the Stackelberg game. In addition, the authors determined the maximum number of devices connected to the network and the upper bound for the ergodic throughput of the cooperative scheme. Also, according to the optimization and analysis results, the suggested two-cell cooperative scheme exhibited more advantages in comparison with the two non-cooperative schemes. Based on the analyses, the suggested cooperative MEC-aided blockchain network enhanced the throughput of the system significantly. In addition, more devices were capable of accessing the blockchain network. Based on the analytical results, the suggested backward induction-based iterative algorithm could reach the games' NE efficiently. According to the numerical results, the suggested backward induction-based iterative algorithm exhibited good stability and fast convergence. Also, the suggested cooperative scheme was capable of serving more devices compared to alternative non-cooperative schemes. As an advantage, the authors in this scheme have done the simulations and evolution stages comprehensively.

**Table 10**: Properties of the cooperative Game theory schemes

| Ref | Utilized techniques | Evaluation Parameters | Architecture | Simulators | Key contributions | Advantages |
|---|---|---|---|---|---|---|
| [122] | | System overhead, Number of iterations, Number of convergences, Ratio, System cost, Number of miners, Reward. | Distributed Computation offloading | N/A | Distributed high-efficiency algorithm based on the limited recovery property of the MCOG. | Distributed algorithm, Acceptable system cost. |
| [123] | Nash equilibrium | Offloading probability, expected latency, Expected payoff. | Distributed Computation offloading | MATLAB | Data sharing architecture among user vehicles and service providers for collaborative computation offloading in vehicular MEC networks based on blockchain. | Byzantine Fault Tolerance, Secure data sharing architecture. |
| [124] | Nash equilibrium | Resource service unit price for BSs and users, Computing demand, User access probability, Block storage strategies, Computing service unit price. | Distributed Computation offloading | CSP | Collaborative MEC-aided blockchain network for PoW mining tasks. | Fast convergence, Acceptable stability. |

### 5.3. Metaheuristic-based schemes

Metaheuristic algorithm operate by repeated evaluations of the objective function, generally without the need for using gradient data [125].

#### 5.3.1. NSGA-III-based schemes

As a genetic algorithm, NSGA-III serves on the basis of reference points. Except for the NSGA-II selection operation, NSGA-III adopts the NSGA-II framework. While NSGA-III employs a collection of reference directions in order to keep the diversity amongst solutions, for the same purpose, a more adaptive scheme is utilized in NSGA-II via its crowding distance operator [126].
In [127], Xu et al. presented BeCome to reduce the energy consumption of ECDs and the task offloading time of ECDs while reaching data integrity and load balance in the Internet of things. They used blockchain technology in edge computing in order to get assured of the integrity of data. Subsequently, they adopted the NSGA-III in order to create some strategies for the balanced allocation of resources. In addition, they used multicriteria decision-making (MCDM) and simple additive weighting (SAW) in order to determine the optimum offloading strategy. Eventually, simulation experiments were conducted to evaluate the performance of BeCome. As a limitation, the authors in this scheme did not generalize BeCome to the real scene of the Internet of things. Also, it was necessary to determine the number of tasks on every single smart device in accordance with the real conditions.
In [128], Xu et al. presented the BCO technique. Given the UEs' computing capacity limitations and hardware constraints, the integration of multi-access edge computing and 5G could support the emerging applications in 5G. From the technical viewpoint, given that blockchain technology was a promising method for decentralized systems, an EC framework was developed on the basis of blockchain in order to decrease the possibility of data losses through the integration of EC and blockchain. Given the unequal distribution of resources, the ENs' operating performance was poorly secured. In addition, the data loss and transmission delays in the course of computation offloading have greatly affected the user QoS. Subsequently, the third version of the non-dominated sorting genetic algorithm, i.e., NSGA-III, was leveraged in order to reach the strategies of balanced offloading. In order to select the optimum offloading strategy, the authors utilized the advantages of SAW and MCDM. Eventually, systematic analyses and experiments were carried out on the comparative experiment in order to verify the efficiency of the BCO technique. As a limitation, the suggested BCO in this scheme was required to be modified in accordance with the concrete requirements of real-world scenarios. In addition, it was necessary to consult more preferences of user QoS in order to transform the suggested technique into a more efficient and practical one.

#### 5.3.2. Other schemes

In general, the strategy in the auction theory is to give/take the selling objects at optimal prices in that both parties remain unsure of the value. The same objects can be defined as various communication and computation services on which the users must agree in the offloading criterion [28].

By employing an approximation ratio of $(1 - \epsilon)$ on the basis of a group-buying mechanism, Xia et al. [129], presented a three-stage auction scheme in order to allocate the resources of edge servers for mobile blockchain purposes. Initially, they divided the miners into some groups, and in order to decide every single group's bid for each edge server, a Vickrey–Clarke–Groves-based auction was presented. AP indicated a group matched with an edge server. They suggested a matching algorithm in the second stage in order to match Access Points and edge servers so as to maximize the edge servers' profits. They allocated the edge server resources to mobile users in the third step for the purpose of mining on the basis of the results obtained in the above stages. The authors showed that the auction scheme presented by them could guarantee individual rationality, budget balance, and truthfulness. They made a comparison between their suggested scheme and HAF, TARCO, and TACD. According to the extensive simulation results, for a community of 1000 miners and concerning social welfare, their scheme outperformed HAF, TACD, SWM, and TARCO mechanisms by 21.84%, 33.78%, 6.69%, and 19.69%, respectively. As a limitation, this scheme required a high-speed and dynamic user movement network medium. In addition, there was a need to design novel online algorithms for resource allocation for a practical and more efficient blockchain system so as to adapt to the complicated communication medium.

By solving an NP-hard multiple-choice multi-dimensional knapsack problem, Li et al. [130], suggested an auction mechanism known as POEM+, which was capable of offloading a mobile user's tasks to edge servers located on heterogeneous edge serves. On the basis of the incentive mechanism, a computation offloading experiment platform was set up in order to decrease the finish time of the tasks. Theoretically, the authors demonstrated that POEM+ could satisfy economic features, including budget balance, individual rationality, computation efficiency, and truthfulness. According to real-world experiments and simulation results, POEM+ reached 130.6% higher utility in comparison with the existing heterogeneous tasks double auction WBD and also 138.77% higher allocation efficiency on average. In addition, they could guarantee the long-term performance of POEM+. As an advantage, a complete review of the simulations and evolution stages has been presented by the authors.

In [131], Liu et al. presented a blockchain-enabled multi-MEC environment so as to transform the problem of utility maximization into a multi-choice knapsack one. They showed the NP-hardness of the problem. In addition, in order to reach a near-optimal solution, a double auction mechanism, known as a long-term auction, was developed for the purpose of inner-dependent task offloading (LAIO). According to the experimental results, the suggested algorithm outperforms the benchmarks (DPESO, PSO, and RTO). Concerning the total utility, LAIO outperformed DPESO, PSO, and RTO by 16.2%, 40.1%, and 53.8%, respectively. In addition, the suggested LAIO algorithm was capable of ensuring long-term computation offloading performance. As an advantage, the authors have provided a comprehensive set of simulations for their scheme.

The efficient task-VM (virtual machine) matching algorithm suggested by Seng et al. [132], was capable of jointly considering the energy consumption and task execution time. In particular, the stability of the suggested matching algorithm was approved. In addition, by the development of a smart matching contract, the authors used the designed task-virtual machine matching algorithm on the blockchain in order to carry out matching on the blockchain without the need for trusted third parties. They carried out extensive simulations in order to validate the efficacy of the presented matching algorithm. According to the extensive simulation results, the decentralized coordination scheme converged to a stable condition very fast and was capable of enhancing performance significantly. As an advantage, the authors in this scheme, have done the simulations and evolution stages comprehensively.

**Table 11**: Properties of the NSGA-based and Auction-based schemes

| **Ref** | **Utilized techniques** | **Evaluation Parameters** | **Architecture** | **simulators** | **Key contributions** | **Advantages** |
|---|---|---|---|---|---|---|
| [127] | non-dominated sorting genetic algorithm III, multicriteria decision-making, simple additive weighting. | Resource utilization, Load balancing rate, Offloading time, Energy consumption | Distributed Computation offloading | N/A | BeCome theory for decreasing task offloading time and the energy consumption of ECDs in the IoT. | Multi-objective optimization. |
| [128] | non-dominated sorting genetic algorithm III multicriteria decision-making simple additive weighting | The utility values, Latency between edge notes, The total latency, The number of occupied edge notes, Average resource utilization, Load balance degree. | Distributed Computation offloading | N/A | BCO scheme for blockchain-based computation migrating. | Effient method, Good presentation of the proposed model. |
| [129] | Matching algorithm | Utility of edge servers, Utility of Aps, Utility of miners, Social welfare. | Decentralized Computation offloading | MATLAB | Mobile blockchain system for maximizing the social welfare for computation offloading. | Efficient scheme, Acceptable complexity. |
| [130] | POEM+ | Total utility, Satisfaction Ratio, Utilization Ratio | Computation offloading | N/A | Truthful incentive mechanism with multiple buyers and numerous sellers. | Efficient computation, Acceptable budget balance. |
| [131] | long-term auction for inner-dependent task offloading | Utilization ratio, Satisfaction ratio, Total utility. | Task offloading | MATLAB | Blockchain-enabled multi-MEC servers offloading, and the NP-hardness of the efficiency maximization difficulty. | Increased the total utility of mobile devices and edge servers, Acceptable complexity. |
| [132] | Matching algorithm | Average task execution time, Average energy consumption, Match rate, Average utility received by each buyer, the average utility received by each seller. | Decentralized Computation offloading | N/A | Decentralized cooperative model to orchestrate MUs and EgSvrs. | Decentralized coordination scheme, Good performance, Stable state. |

## 6. Discussion

Figure 12 illustrates the distribution of different types of offloading schemes utilized in MEC environments, with a clear dominance of Machine Learning-based approaches, which constitute 61% of the schemes analyzed. Game Theory-based methods are also significant, accounting for 29%, demonstrating their robust strategic decision-making capabilities in MEC. Auction-based models hold a smaller portion at 6%, suggesting a niche application in scenarios that may benefit from competitive bidding processes. Finally, Metaheuristic-based methods make up 4% of the schemes, indicating a relatively limited, yet specialized, application possibly due to their complexity and computational requirements. This distribution highlights the prevailing trend and potential areas for further research in the offloading strategies for MEC.

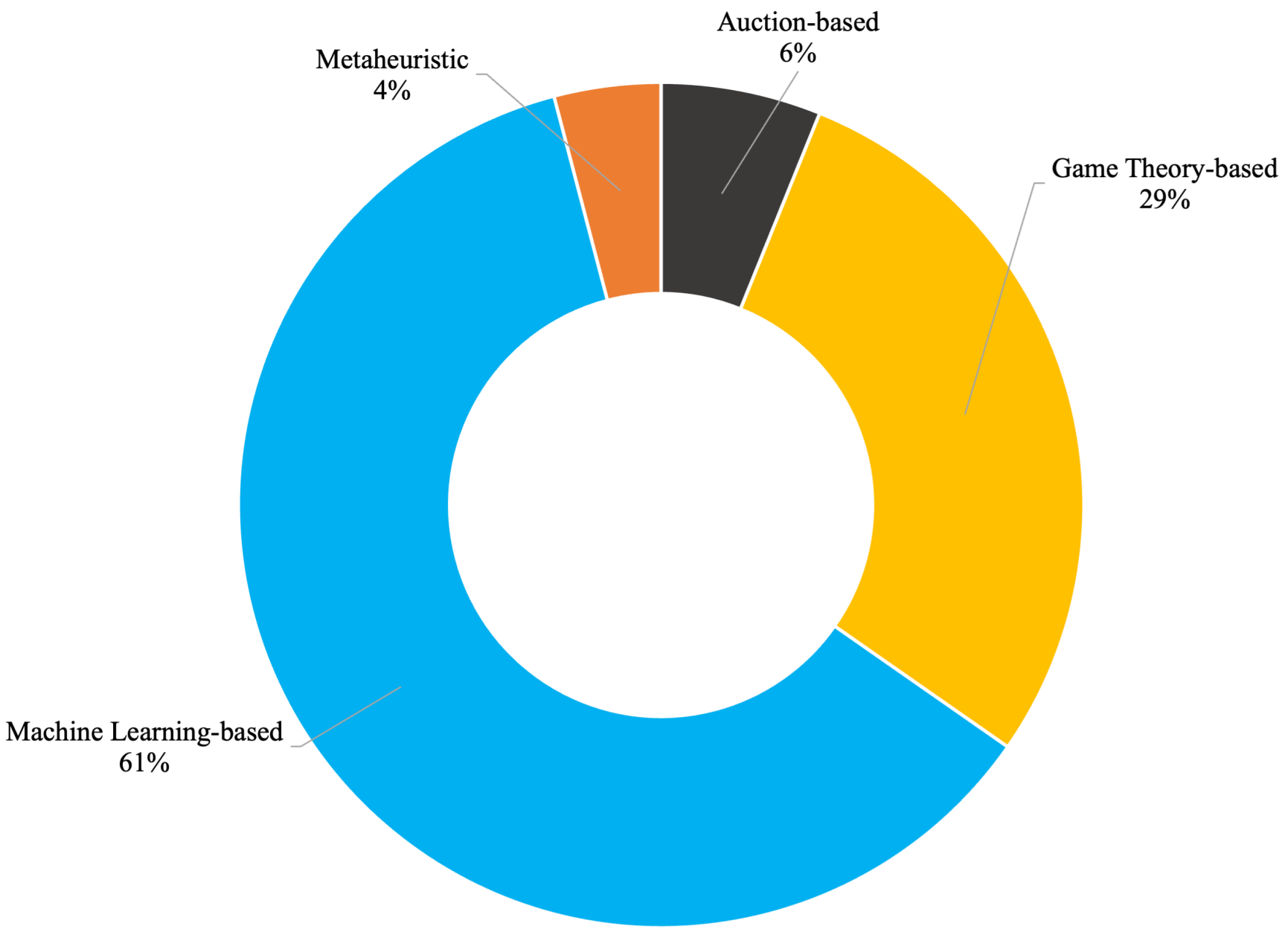

**Fig. 12.** Classification of blockchain-based offloading schemes in MECs

Figure 13 depicting the proportion of various offloading types within a certain context, presumably MEC. The largest section, at 54%, is dedicated to Computation offloading, which is likely indicative of the emphasis on processing tasks being transferred from local devices to the edge servers. Task offloading follows with 38%, showing a significant reliance on delegating specific tasks to improve efficiency and possibly reduce latency or energy consumption on the client devices. Lastly, Data offloading is represented as the smallest segment at 8%, suggesting that while still important, it's less prioritized compared to the other types of offloading. This may be due to inherent challenges or lower perceived benefits of offloading data as opposed to computational tasks. Overall, the chart emphasizes that computation and task offloading are key strategies employed in the area of study, with data offloading playing a less prominent, yet still crucial role.

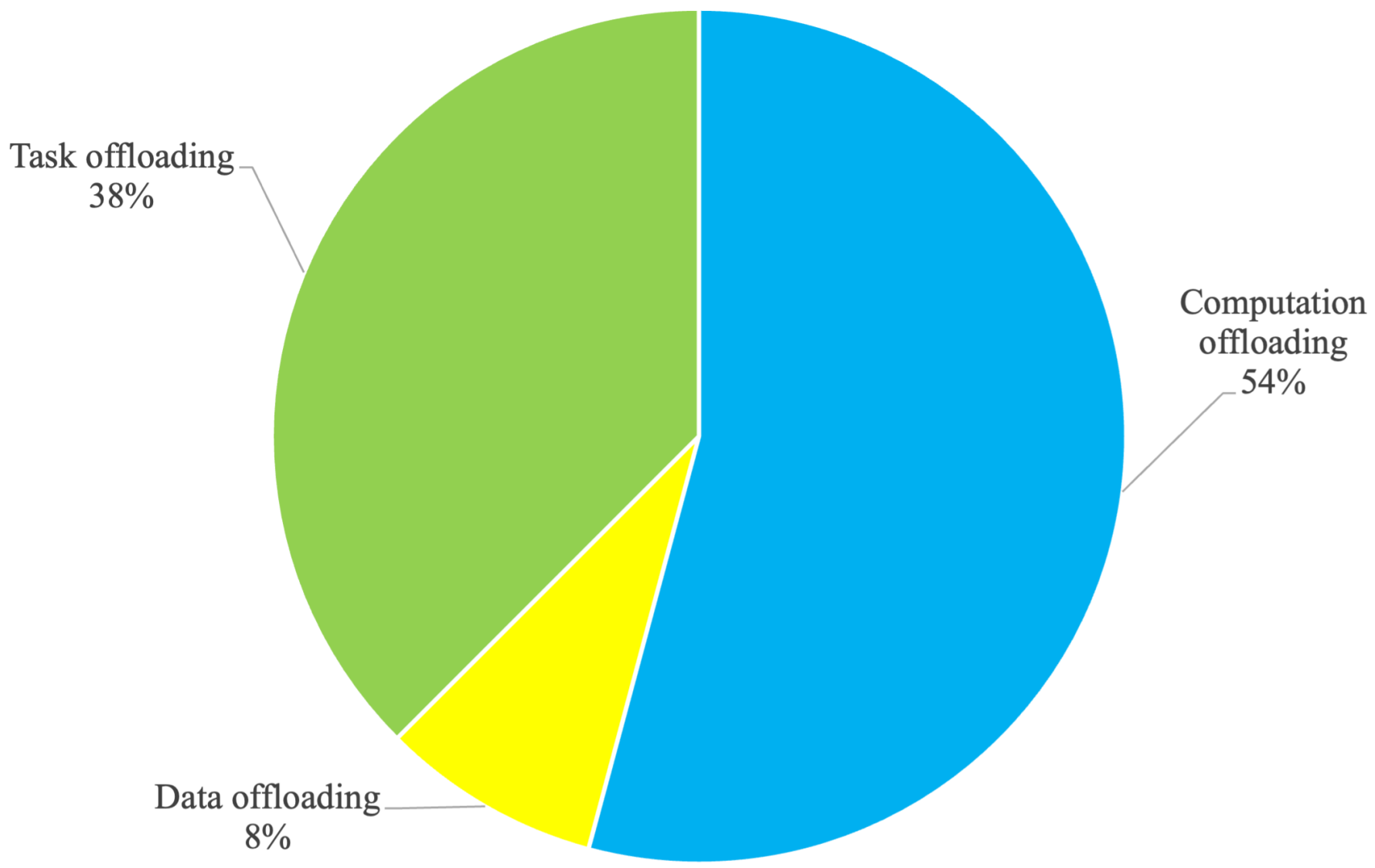

**Fig. 13.** Offloading types

Figure 14 displays a clear preference for architectural choices in the schemes analyzed, with a striking 42 schemes employing a decentralized architecture, underscoring its prominence in the field. In stark contrast, only 3 schemes opted for a centralized architecture, suggesting possible limitations or specific use cases for such an approach. Additionally, 4 schemes are marked as not applicable (NA), which could denote a divergence from the traditional centralized/decentralized dichotomy or perhaps insufficient data regarding their architectural framework. The data overwhelmingly suggests that decentralized architecture is the preferred model in this area of study, reflecting its suitability for enhancing system robustness and distributive efficiency.

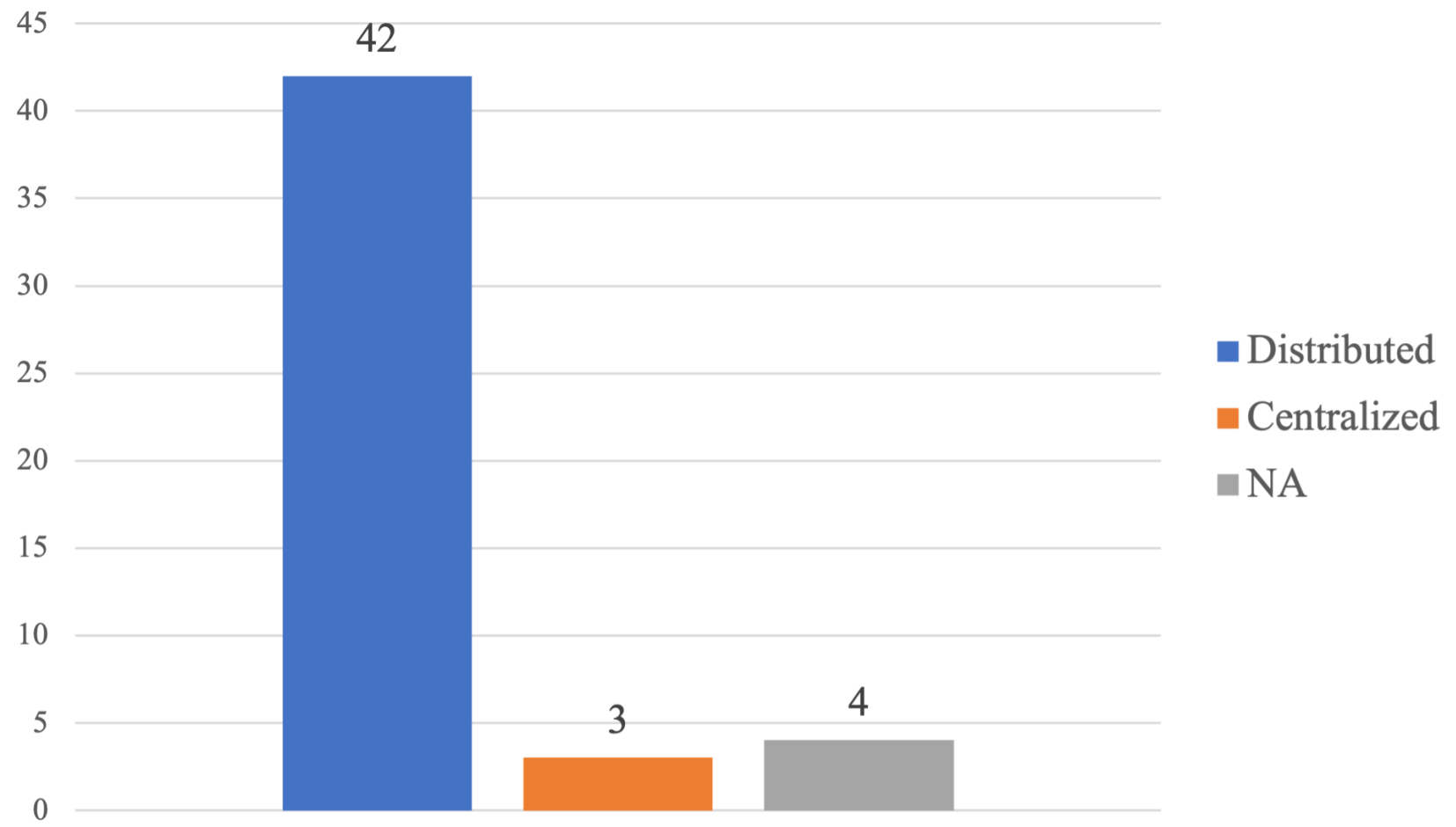

**Fig. 14.** System model

Figure 15 depicts the programming languages and tools utilized in the analyzed schemes. A majority, 55%, did not specify their tools (NA), indicating a gap in reporting or a diversity of unspecified tools. Python is the preferred language for 33% of the schemes, highlighting its widespread adoption due to its versatility and robust libraries. Matlab is used in 6%, followed by both NS3 and Opnet at 2% each, which are network simulation tools. CSP, another programming tool, is also employed by 2% of the schemes.

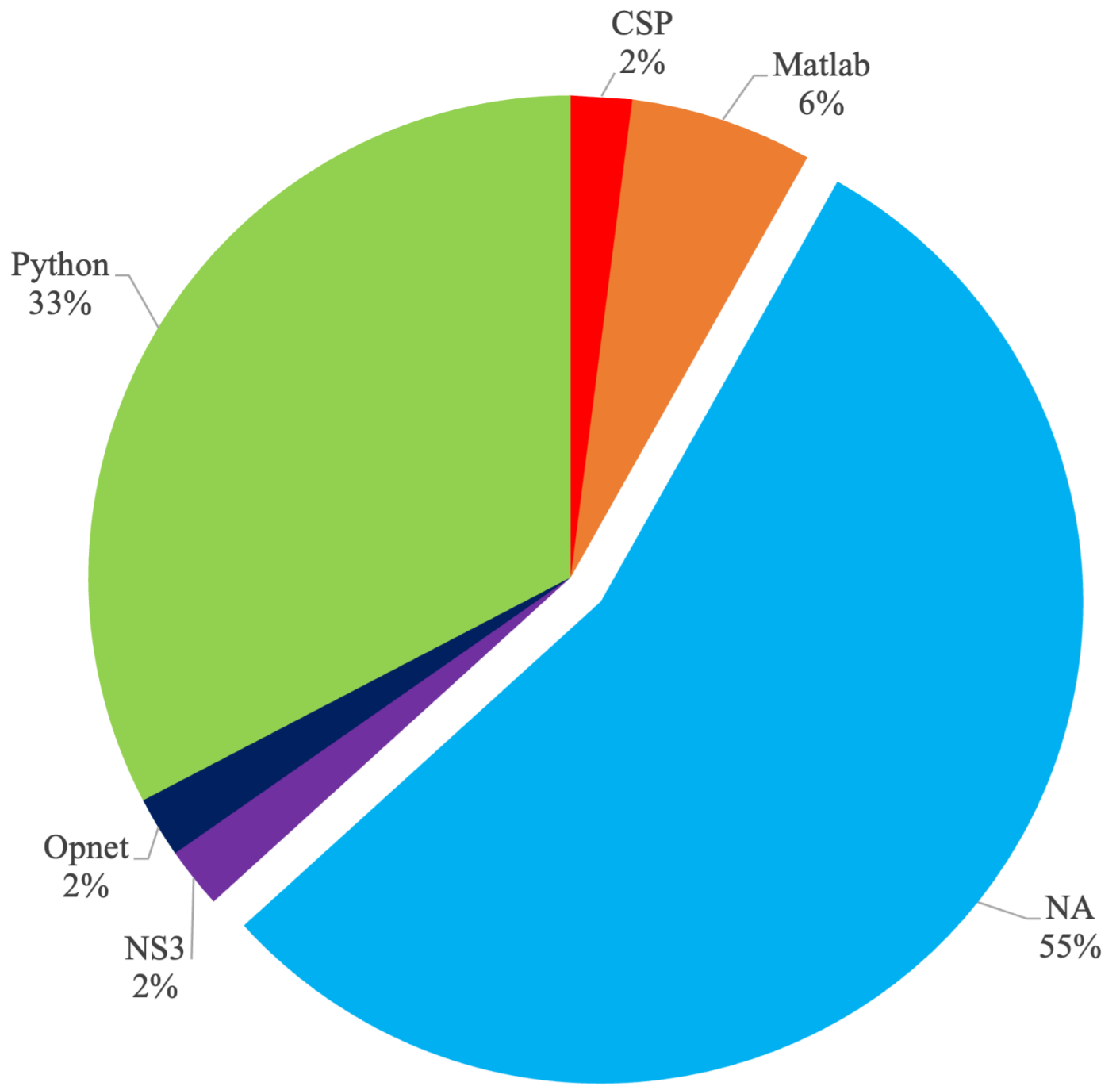

**Fig. 15.** Applied simulators and tools

Figure 16 exhibits a bar chart detailing the utilization of different evaluation metrics across a number of schemes. Reward is the most frequently measured metric, with 18 schemes assessing it, suggesting a focus on the benefits or gains achieved through the use of the schemes. Throughput is also a significant metric, used by 14 schemes, indicating the importance of data processing rates in these analyses. Latency is evaluated by 11 schemes, emphasizing the value placed on the responsiveness of the systems. Other metrics such as Energy Consumption, Time Cost, and System Utility show varied levels of attention, pointing to a multifaceted approach to performance evaluation in the field. Metrics like Loss, Accuracy are considered in the least number of schemes, reflecting their specialized or less prioritized status in this context.

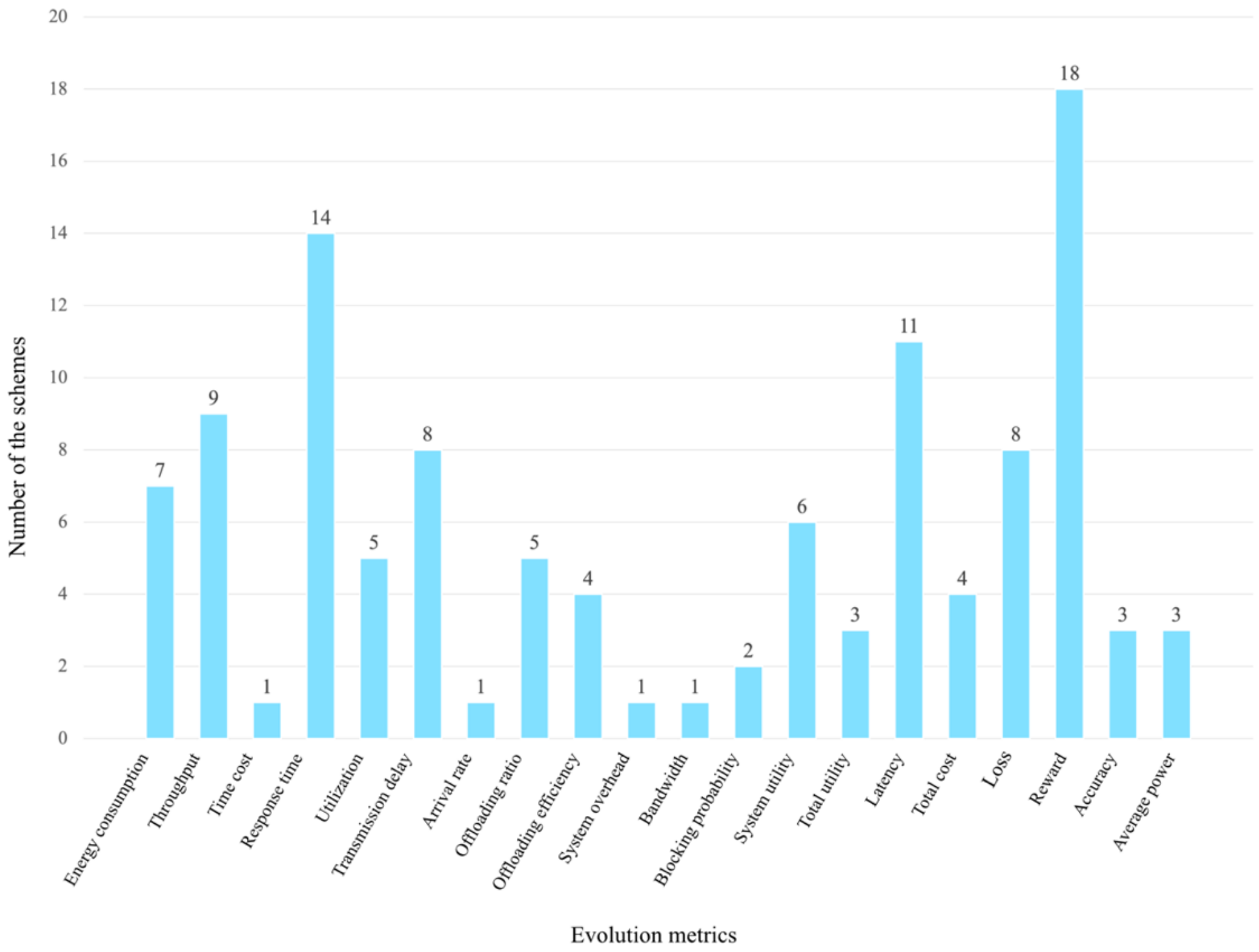

**Fig. 16.** Applied metrics

Figure 17 is a horizontal bar chart presenting the number of schemes utilizing different algorithm types. DRL is the most applied approach with 15 schemes using this technique, indicating its significance and popularity in the field. MDP are also prominent, employed in 7 schemes, reflecting their importance in decision-making scenarios. Other methods such as A3C, FL, DL, and various forms of Q-Learning are less frequently used, ranging from 2 to 5 schemes. HRL and HDL are the least utilized, with only 1 scheme each, suggesting they are highly specialized or new in this area of research.

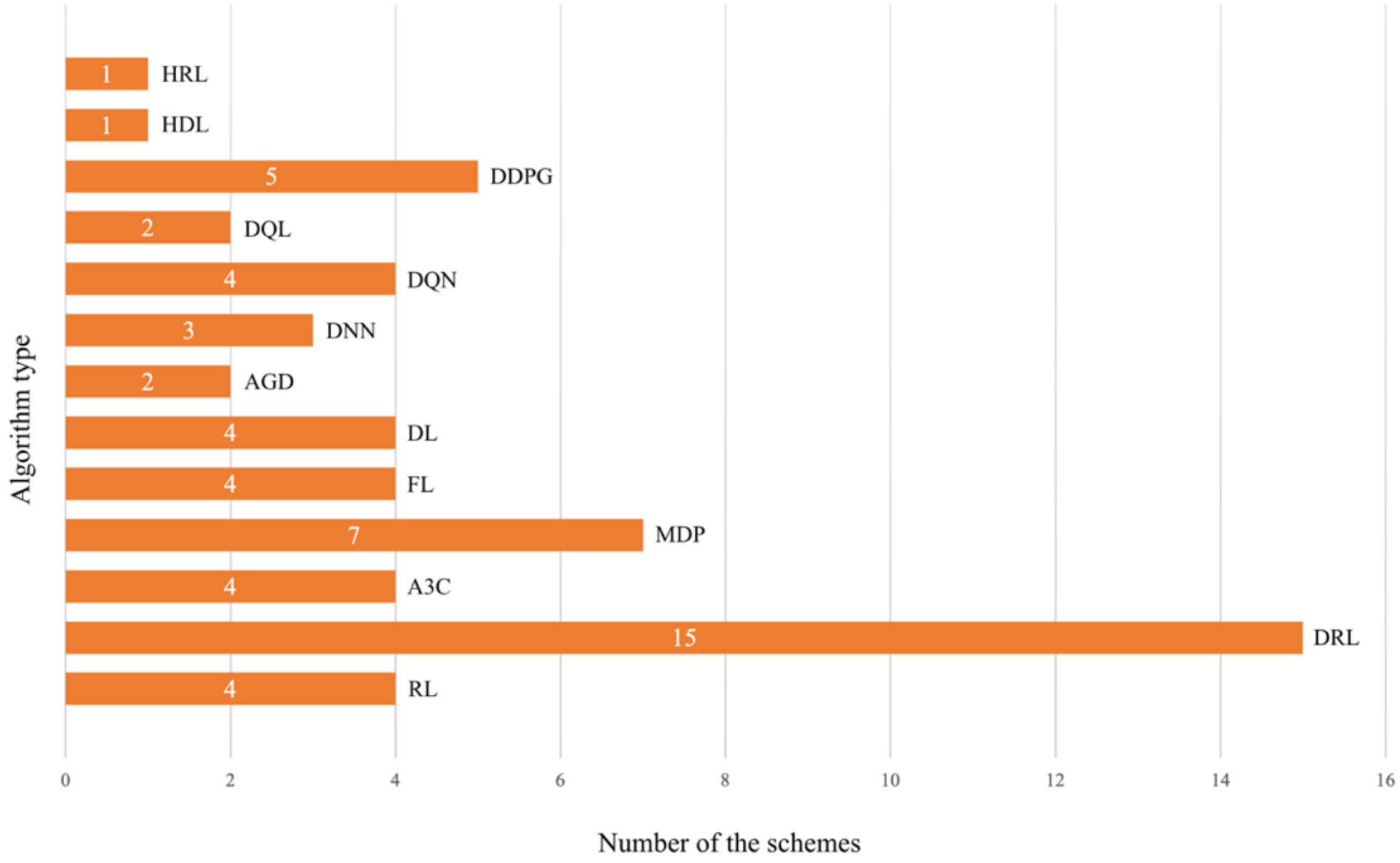

**Fig. 17.** The distribution of different techniques in machine learning-based schemes

Fig. 18a and Fig .18b offer insights into the distribution of game theory-based schemes. In Fig. 18a, a significant majority of schemes, 77%, are categorized as non-cooperative, indicating a prevalence of strategies where individual agents act independently without collaboration. The remaining 23% of schemes utilize a cooperative-based approach, suggesting scenarios where entities work together to achieve a common goal. Fig. 18b. further categorizes these game theory-based schemes. The Stackelberg model is the most prominent, employed by 46% of the schemes, reflecting its utility in scenarios where a leader-follower dynamic is advantageous. Nash equilibrium concepts are used in 31% of the schemes, illustrating their importance in finding stable strategies among competing agents. Potential games account for 15%, indicative of situations where outcomes can be improved through coordinated changes in strategy. Lastly, stochastic games are the least represented at 8%, pointing to their specific application in environments with probabilistic elements.

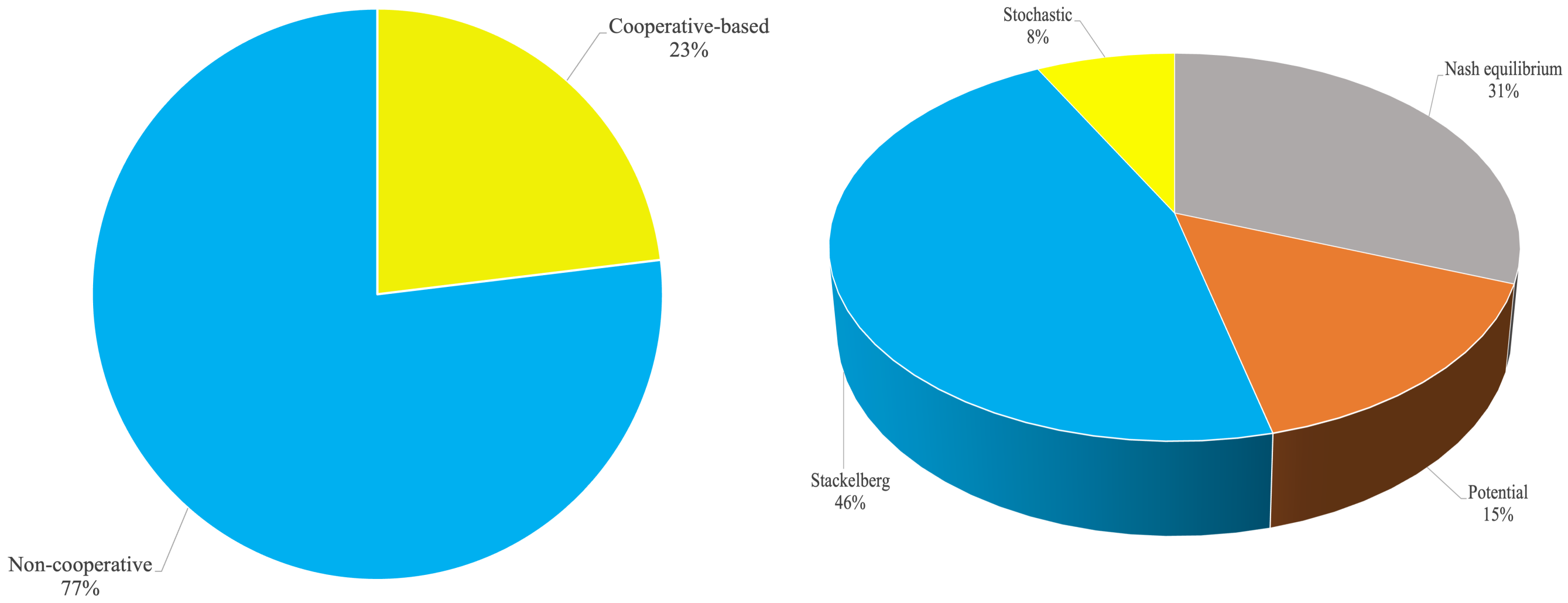

**18.a.** **18.b**

**Fig. 18a, b.** The distribution of different techniques in game theory-based schemes.

## 7. Future Directions and Challenges

The exploration of blockchain-enabled offloading in MEC presents numerous opportunities for advancement and innovation. Yet, this burgeoning field faces multiple challenges that need to be addressed. Here, we outline future research directions to enhance the efficacy and applicability of MEC offloading schemes.

**7.1. Integration of Emerging Technologies:** Research should explore the integration of nascent technologies such as quantum computing and artificial intelligence to bolster MEC offloading schemes' security and efficiency.

**7.2. Multi-access Edge Computing (MEC) Server Enhancement:** Future studies must focus on augmenting the capacity and capabilities of MEC servers to handle increasingly complex offloading tasks with minimal latency.

**7.3. Enhancement of Blockchain Protocols:** To address scalability and speed concerns, the development of new or improved blockchain protocols that are tailor-made for MEC environments is crucial.

**7.4. Granular Data Privacy Controls:** Investigating sophisticated privacy-preserving mechanisms that offer users granular control over their data during the offloading process is an imperative research area.

**7.5. Offloading for Ultra-Reliable Low-Latency Communication (URLLC):** Given the rise of 5G, offloading solutions that cater to URLLC requirements should be a primary focus to support emerging real-time applications.

**7.6. Energy Harvesting Techniques:** The incorporation of energy harvesting techniques into offloading decisions can prolong device lifespans and promote sustainable MEC practices.

**7.7. Inter-Blockchain Communication:** Facilitating seamless communication and offloading across different blockchain platforms to enhance interoperability and flexibility in MEC settings is a key challenge.

**7.8. Holistic Economic Models:** Developing comprehensive economic models that consider the entire ecosystem, including MEC providers, users, and blockchain platforms, is vital for creating balanced offloading schemes.

**7.9. Autonomic Offloading Mechanisms:** Creating self-managing offloading mechanisms that autonomously adapt to changing network conditions and user demands is an important direction for research.

**7.10. Offloading in Edge AI Applications:** Investigating offloading's role in edge AI applications, including federated learning, can expand the horizon for intelligent edge computing solutions.

**7.11. Cross-Layer Optimization:** Research should aim at cross-layer optimization strategies that consider the offloading impact from the physical layer up to the application layer.

**7.12. Resilience to Adversarial Machine Learning:** Strengthening offloading mechanisms against adversarial machine learning attacks is a burgeoning area that warrants attention.

**7.13. Edge Cloud Synergies:** Exploring deeper synergies between edge computing and cloud infrastructures to leverage the strengths of both in offloading strategies is a valuable pursuit.

**7.14. User-Centric Offloading Frameworks:** Future research should aim to develop user-centric offloading frameworks that prioritize user experience in terms of speed, cost, and service quality.

**7.15. Offloading for Internet of Vehicles (IoV):** Delving into offloading solutions specifically tailored for the IoV, which is poised to be a major generator of data in need of computation, is a promising research area.

Tackling these challenges and exploring these avenues will be crucial for advancing MEC offloading schemes, ensuring they meet the needs of an increasingly connected and data-driven world.

## 8. Conclusion

In conclusion, this review has critically examined the integration of blockchain technology within MEC offloading schemes. Our comprehensive survey reveals a burgeoning field ripe with potential, characterized by a predilection for decentralized approaches and an emerging focus on diverse methodologies. The challenges identified necessitate innovative solutions to enhance security, efficiency, and adaptability. Future research directions are abundant, promising a vibrant landscape for advancements that will shape the next generation of MEC offloading. As the IoT ecosystem evolves, these insights will be pivotal in steering the trajectory of this technologically pivotal domain.

**References**

1. Laghari, A.A., et al., *A review and state of art of Internet of Things (IoT).* 2021: p. 1-19.
2. Peng, S.-L., S. Pal, and L. Huang, *Principles of internet of things (IoT) ecosystem: Insight paradigm*. 2020: Springer.
3. Leminen, S., et al., *Industrial internet of things business models in the machine-to-machine context.* 2020. **84**: p. 298-311.
4. Zhao, T., et al., *DRL-Based Secure Video Offloading in MEC-Enabled IoT Networks.* 2022. **9**(19): p. 18710-18724.
5. Zhang, X., et al., *An efficient computation offloading and resource allocation algorithm in RIS empowered MEC.* 2023. **197**: p. 113-123.
6. Saidi, K. and D. Bardou, *Task scheduling and VM placement to resource allocation in cloud computing: challenges and opportunities.* Cluster Computing, 2023. **26**(5): p. 3069-3087.
7. Wu, H., et al., *EEDTO: an energy-efficient dynamic task offloading algorithm for blockchain-enabled IoT-edge-cloud orchestrated computing.* 2020. **8**(4): p. 2163-2176.
8. Kuang, F., et al., *Multi-workflow scheduling and resource provisioning in Mobile Edge Computing using opposition-based Marine-Predator Algorithm.* 2022. **87**: p. 101715.
9. Moghaddasi, K. and S. Rajabi. *Learning at the Edge: Mobile Edge Computing and Reinforcement Learning for Enhanced Web Application Performance*. in *2023 9th International Conference on Web Research (ICWR)*. 2023.
10. Letaief, K.B., et al., *Edge artificial intelligence for 6G: Vision, enabling technologies, and applications.* 2021. **40**(1): p. 5-36.
11. Bréhon–Grataloup, L., R. Kacimi, and A.-L.J.C.N. Beylot, *Mobile edge computing for V2X architectures and applications: A survey.* 2022. **206**: p. 108797.
12. Lu, J., et al., *Analytical offloading design for mobile edge computing-based smart internet of vehicle.* 2022. **2022**(1): p. 44.

13. Akhlaqi, M.Y. and Z.B. Mohd Hanapi, *Task offloading paradigm in mobile edge computing-current issues, adopted approaches, and future directions.* Journal of Network and Computer Applications, 2023. **212**: p. 103568.
14. Kumaran, K. and E.J.M.S. Sasikala, *Computational access point selection based on resource allocation optimization to reduce the edge computing latency.* 2022. **24**: p. 100444.
15. Moghaddasi, K., S. Rajabi, and F.S. Gharehchopogh, *An enhanced asynchronous advantage actor-critic-based algorithm for performance optimization in mobile edge computing -enabled internet of vehicles networks.* Peer-to-Peer Networking and Applications, 2024.
16. Moghaddasi, K., et al., *An Energy-Efficient Data Offloading Strategy for 5G-Enabled Vehicular Edge Computing Networks Using Double Deep Q-Network.* Wireless Personal Communications, 2023. **133**(3): p. 2019-2064.
17. Ma, H., et al., *Video data offloading techniques in Mobile Edge Computing: A survey.* Physical Communication, 2023: p. 102261.
18. Moghaddasi, K. and S. Rajabi. *Double Deep Q-Learning Networks for Energy-Efficient IoT Task Offloading in D2D MEC Environments*. in *2023 7th International Conference on Internet of Things and Applications (IoT)*. 2023.
19. Mach, P., Z.J.I.c.s. Becvar, and tutorials, *Mobile edge computing: A survey on architecture and computation offloading.* 2017. **19**(3): p. 1628-1656.
20. Chai, F., et al., *Joint multi-task offloading and resource allocation for mobile edge computing systems in satellite iot.* IEEE Transactions on Vehicular Technology, 2023.
21. Maray, M. and J.J.M.I.S. Shuja, *Computation offloading in mobile cloud computing and mobile edge computing: survey, taxonomy, and open issues.* 2022. **2022**.
22. Moghaddasi, K., S. Rajabi, and F.S. Gharehchopogh, *Multi-Objective Secure Task Offloading Strategy for Blockchain-Enabled IoV-MEC Systems: A Double Deep Q-Network Approach.* IEEE Access, 2024. **12**: p. 3437-3463.
23. Andoni, M., et al., *Blockchain technology in the energy sector: A systematic review of challenges and opportunities.* 2019. **100**: p. 143-174.
24. Moghaddasi, K. and M. Masdari, *Blockchain-driven optimization of IoT in mobile edge computing environment with deep reinforcement learning and multi-criteria decision-making techniques.* Cluster Computing, 2023.
25. Xiong, Z., et al., *When mobile blockchain meets edge computing. arXiv 2017.*
26. Masuduzzaman, M., et al., *UAV-based MEC-assisted automated traffic management scheme using blockchain.* 2022. **134**: p. 256-270.
27. Zhao, N., H. Wu, and Y.J.I.i.o.t.j. Chen, *Coalition game-based computation resource allocation for wireless blockchain networks.* 2019. **6**(5): p. 8507-8518.
28. Shahidinejad, A.S.A. and M. Ghobaei-Arani, *A review on the computation offloading approaches in mobile edge computing: A game-theoretic perspective.*
29. Shakarami, A., M. Ghobaei-Arani, and A.J.C.N. Shahidinejad, *A survey on the computation offloading approaches in mobile edge computing: A machine learning-based perspective.* 2020. **182**: p. 107496.
30. Liao, Z., et al., *Blockchain on security and forensics management in edge computing for IoT: A comprehensive survey.* 2021. **19**(2): p. 1159-1175.
31. Feng, C., et al., *Computation offloading in mobile edge computing networks: A survey.* 2022: p. 103366.
32. Xue, H., et al., *Integration of blockchain and edge computing in internet of things: A survey.* 2022.
33. Davidson, S., P. De Filippi, and J.J.A.a.S. Potts, *Economics of blockchain.* 2016.
34. Sullivan, F., E. Mucke, and M.J.n.O. Di Pierro, *SECTION TITLE COMPUTING PRESCRIPTIONS What Is the Blockchain?* 2017.
35. Laali, K.K., et al., *Novel fluorinated curcuminoids and their pyrazole and isoxazole derivatives: Synthesis, structural studies, Computational/Docking and in-vitro bioassay.* 2018. **206**: p. 82-98.
36. Guo, H., X.J.B.r. Yu, and applications, *A Survey on Blockchain Technology and its security.* 2022. **3**(2): p. 100067.
37. Banerjee, M., J. Lee, and K.J.U.h.w.s.c.s.a.p.S. Choo, *A blockchain future to Internet of Things security: A position paper, Digital Communications and Networks (2017).*
38. Neudecker, T., H.J.I.C.S. Hartenstein, and Tutorials, *Network layer aspects of permissionless blockchains.* 2018. **21**(1): p. 838-857.
39. Wang, X., et al., *Survey on blockchain for Internet of Things.* 2019. **136**: p. 10-29.
40. Yaqoob, S., et al., *Use of blockchain in healthcare: a systematic literature review.* 2019. **10**(5).
41. Ram, A., W. Maroun, and R.J.M.A.R. Garnett, *Accounting for the Bitcoin: accountability, neoliberalism and a correspondence analysis.* 2016. **24**(1): p. 2-35.
42. Tuwiner, J.J.O.v.u.h.w.b.c.m.p., zuletzt geprüft am, *Bitcoin mining pools.* 2019. **13**: p. 2019.
43. Wong, J.I., *China's Bitmain dominates bitcoin mining. Now it wants to cash in on artificial intelligence*. 2017, Quartz.
44. Zheng, Z., et al. *An overview of blockchain technology: Architecture, consensus, and future trends*. in *2017 IEEE international congress on big data (BigData congress)*. 2017. Ieee.
45. Condoluci, M., et al., *Enabling the IoT machine age with 5G: Machine-type multicast services for innovative real-time applications.* 2016. **4**: p. 5555-5569.
46. Alao, O. and P. Cuffe. *A Taxonomy of the Risks and Challenges of Embracing Blockchain Smart Contracts in Facilitating Renewable Electricity Transactions*. in *2022 IEEE PES/IAS PowerAfrica*. 2022. IEEE.
47. Ponce, E., J. Mula, and D. Peidro, *A Blockchain Applications Overview*, in *Ensuring Sustainability: New Challenges for Organizational Engineering*. 2022, Springer. p. 85-95.
48. AsteT, T.J.C., *DiMatteoT. Blockchaintechnologies: Theforeseeableimpactonsocietyandindustry.* 2017. **50**(9): p. 18.
49. Bhutta, M.N.M., et al., *A survey on blockchain technology: Evolution, architecture and security.* 2021. **9**: p. 61048-61073.
50. Ferrag, M.A., et al., *Blockchain technologies for the internet of things: Research issues and challenges.* 2018. **6**(2): p. 2188-2204.

51. Saputhanthri, A., C. De Alwis, and M.J.I.A. Liyanage, *Survey on Blockchain-Based IoT Payment and Marketplaces.* 2022. **10**: p. 103411-103437.
52. Alzhrani, F.E., K.A. Saeedi, and L.J.I.A. Zhao, *A Taxonomy for Characterizing Blockchain Systems.* 2022. **10**: p. 110568-110589.
53. Huang, H., et al., *Fusion of building information modeling and blockchain for metaverse: a survey.* 2022. **3**: p. 195-207.
54. Yang, R., et al., *Integrated blockchain and edge computing systems: A survey, some research issues and challenges.* 2019. **21**(2): p. 1508-1532.
55. Xie, J., et al., *A survey of blockchain technology applied to smart cities: Research issues and challenges.* 2019. **21**(3): p. 2794-2830.
56. Zhu, C., et al., *Blockchain-enabled federated learning for UAV edge computing network: Issues and solutions.* 2022. **10**: p. 56591-56610.
57. Chukwu, E. and L.J.I.A. Garg, *A systematic review of blockchain in healthcare: frameworks, prototypes, and implementations.* 2020. **8**: p. 21196-21214.
58. Tapscott, D. and A. Tapscott, *Blockchain revolution: how the technology behind bitcoin is changing money, business, and the world*. 2016: Penguin.
59. Ali, O., et al., *A comparative study: Blockchain technology utilization benefits, challenges and functionalities.* 2021. **9**: p. 12730-12749.
60. Tandon, A., et al., *Blockchain applications in management: A bibliometric analysis and literature review.* 2021. **166**: p. 120649.
61. Ghobaei-Arani, M., A. Souri, and A.A.J.J.o.G.C. Rahmanian, *Resource management approaches in fog computing: a comprehensive review.* 2020. **18**(1): p. 1-42.
62. Chen, M.-H., M. Dong, and B.J.I.T.o.M.C. Liang, *Resource sharing of a computing access point for multi-user mobile cloud offloading with delay constraints.* 2018. **17**(12): p. 2868-2881.
63. Tang, M. and V.W.J.I.T.o.M.C. Wong, *Deep reinforcement learning for task offloading in mobile edge computing systems.* 2020. **21**(6): p. 1985-1997.
64. Ghobaei-Arani, M., et al., *An autonomous resource provisioning framework for massively multiplayer online games in cloud environment.* 2019. **142**: p. 76-97.
65. Zhang, Z., et al., *A Clustering Offloading Decision Method for Edge Computing Tasks Based on Deep Reinforcement Learning.* 2022: p. 1-24.
66. Huang, L., et al., *Distributed deep learning-based offloading for mobile edge computing networks.* 2018: p. 1-8.
67. Duato, J., et al. *rCUDA: Reducing the number of GPU-based accelerators in high performance clusters*. in *2010 International Conference on High Performance Computing & Simulation*. 2010. IEEE.
68. Chen, J., et al., *Multitask offloading strategy optimization based on directed acyclic graphs for edge computing.* 2021. **9**(12): p. 9367-9378.
69. Lee, H., J. Choi, and D.J.I.T.o.W.C. Hong, *Resource configuration for full-duplex-aided multiple-access edge computation offloading.* 2021. **21**(4): p. 2799-2812.
70. Elgendy, I.A., et al., *An efficient and secured framework for mobile cloud computing.* 2018. **9**(1): p. 79-87.
71. Akherfi, K., et al., *Mobile cloud computing for computation offloading: Issues and challenges.* 2018. **14**(1): p. 1-16.
72. Wu, G., et al., *Privacy-preserving offloading scheme in multi-access mobile edge computing based on MADRL.* Journal of Parallel and Distributed Computing, 2024. **183**: p. 104775.
73. Joshi, A.V., *Machine learning and artificial intelligence.* 2020.
74. Penmetcha, M. and B.-C.J.I.A. Min, *A deep reinforcement learning-based dynamic computational offloading method for cloud robotics.* 2021. **9**: p. 60265-60279.
75. Liu, Y., et al., *Blockchain-based task offloading for edge computing on low-quality data via distributed learning in the internet of energy.* 2021. **40**(2): p. 657-676.
76. Yao, S., et al., *Blockchain-empowered collaborative task offloading for cloud-edge-device computing.* 2022. **40**(12): p. 3485-3500.
77. Nguyen, D.C., et al., *Privacy-preserved task offloading in mobile blockchain with deep reinforcement learning.* 2020. **17**(4): p. 2536-2549.
78. Nguyen, D.C., et al. *Deep reinforcement learning for collaborative offloading in heterogeneous edge networks*. in *2021 IEEE/ACM 21st International Symposium on Cluster, Cloud and Internet Computing (CCGrid)*. 2021. IEEE.
79. Zhang, H., et al., *Mobility management for blockchain-based ultra-dense edge computing: A deep reinforcement learning approach.* 2021. **20**(11): p. 7346-7359.
80. Alam, T., et al., *Deep reinforcement learning approach for computation offloading in blockchain-enabled communications systems.* 2022: p. 1-14.
81. Nguyen, D.C., et al., *Secure computation offloading in blockchain based IoT networks with deep reinforcement learning.* 2021. **8**(4): p. 3192-3208.
82. Alshaikhi, A. and D.B. Rawat. *Energy Efficient Dynamic Task Offloading for Blockchain-enabled Virtual Wireless Networks*. in *2021 IEEE 12th Annual Ubiquitous Computing, Electronics & Mobile Communication Conference (UEMCON)*. 2021. IEEE.
83. Qiu, X., et al., *Online deep reinforcement learning for computation offloading in blockchain-empowered mobile edge computing.* 2019. **68**(8): p. 8050-8062.
84. Li, K., et al., *Entropy-based Reinforcement Learning for computation offloading service in software-defined multi-access edge computing.* 2022. **136**: p. 241-251.
85. Zhou, Y., et al. *Blockchain-based Trustworthy Service Caching and Task Offloading for Intelligent Edge Computing*. in *2021 IEEE Global Communications Conference (GLOBECOM)*. 2021. IEEE.
86. Samy, A., et al., *Secure task offloading in blockchain-enabled mobile edge computing with deep reinforcement learning.* 2022.

87. Nguyen, D.C., et al., *Intelligent blockchain-based edge computing via deep reinforcement learning: Solutions and challenges.* 2022. **36**(6): p. 12-19.
88. Sellami, B., A. Hakiri, and S.B.J.F.G.C.S. Yahia, *Deep Reinforcement Learning for energy-aware task offloading in join SDN-Blockchain 5G massive IoT edge network.* 2022. **137**: p. 363-379.
89. Masdari, M. and H.J.C. Khezri, *Efficient offloading schemes using Markovian models: a literature review.* 2020. **102**(7): p. 1673-1716.
90. Li, M., et al., *UAV-assisted data transmission in blockchain-enabled M2M communications with mobile edge computing.* 2020. **34**(6): p. 242-249.
91. Lang, P., et al. *Mobility-Aware Computation Offloading and Blockchain-based Handover in Vehicular Edge Computing Networks*. in *2022 IEEE 25th International Conference on Intelligent Transportation Systems (ITSC)*. 2022. IEEE.
92. Zhang, R., et al., *Deep reinforcement learning (DRL)-based device-to-device (D2D) caching with blockchain and mobile edge computing.* 2020. **19**(10): p. 6469-6485.
93. Li, Z., et al., *NOMA-enabled cooperative computation offloading for blockchain-empowered Internet of Things: A learning approach.* 2020. **8**(4): p. 2364-2378.
94. Mathew, A., et al., *Deep learning techniques: an overview.* 2021: p. 599-608.
95. Chu, C.-H.J.W.N., *Task offloading based on deep learning for blockchain in mobile edge computing.* 2021. **27**(1): p. 117-127.
96. Asheralieva, A. and D.J.I.I.o.T.J. Niyato, *Distributed dynamic resource management and pricing in the IoT systems with blockchain-as-a-service and UAV-enabled mobile edge computing.* 2019. **7**(3): p. 1974-1993.
97. Wang, L., et al. *Optimal Energy Efficiency for Multi-MEC and Blockchain Empowered IoT: a Deep Learning Approach*. in *ICC 2021-IEEE International Conference on Communications*. 2021. IEEE.
98. Nguyen, D.C., et al., *Federated learning for internet of things: A comprehensive survey.* 2021. **23**(3): p. 1622-1658.
99. Wang, J., et al. *Privacy-Preserved Computation Offloading Scheme in 5G enabled IoT Based on Smart blockchain*. in *2022 IEEE 10th International Conference on Computer Science and Network Technology (ICCSNT)*. 2022. IEEE.
100. Zhao, Y., et al., *Privacy-preserving blockchain-based federated learning for IoT devices.* 2020. **8**(3): p. 1817-1829.
101. Alghamdi, A., et al., *Blockchain Empowered Federated Learning Ecosystem for Securing Consumer IoT Features Analysis.* 2022. **22**(18): p. 6786.
102. Wang, W., et al., *Privacy protection federated learning system based on blockchain and edge computing in mobile crowdsourcing.* 2022. **215**: p. 109206.
103. Lillicrap, T.P., et al., *Continuous control with deep reinforcement learning.* 2015.
104. Hu, Z., et al., *Joint Optimization for Mobile Edge Computing-Enabled Blockchain Systems: A Deep Reinforcement Learning Approach.* 2022. **22**(9): p. 3217.
105. Nguyen, D., et al., *Cooperative task offloading and block mining in blockchain-based edge computing with multi-agent deep reinforcement learning.* 2021.
106. Qi, J., et al., *Research on an intelligent computing offloading model for the Internet of Vehicles based on Blockchain.* 2022.
107. Hernandez-Leal, P., et al., *A survey and critique of multiagent deep reinforcement learning.* 2019. **33**(6): p. 750-797.
108. Feng, J., et al., *Cooperative computation offloading and resource allocation for blockchain-enabled mobile edge computing: A deep reinforcement learning approach.* 2020. **7**(7): p. 6214-6228.
109. Xu, S., et al., *Deep reinforcement learning assisted edge-terminal collaborative offloading algorithm of blockchain computing tasks for energy Internet.* 2021. **131**: p. 107022.
110. Liu, Z., et al., *A game theory based CTU-level bit allocation scheme for HEVC region of interest coding.* 2020. **30**: p. 794-805.
111. Mkiramweni, M.E., et al., *A survey of game theory in unmanned aerial vehicles communications.* 2019. **21**(4): p. 3386-3416.
112. Khan, M.A. and Y.J.H.o.g.t.w.e.a. Sun, *Non-cooperative games with many players.* 2002. **3**: p. 1761-1808.
113. Xiao, T., et al., *Consortium Blockchain-Based Computation Offloading Using Mobile Edge Platoon Cloud in Internet of Vehicles.* 2022. **23**(10): p. 17769-17783.
114. Liu, W., et al. *A distributed game theoretic approach for blockchain-based offloading strategy*. in *ICC 2020-2020 IEEE International Conference on Communications (ICC)*. 2020. IEEE.
115. Liu, M., et al., *Computation offloading and content caching in wireless blockchain networks with mobile edge computing.* 2018. **67**(11): p. 11008-11021.
116. Zhang, Z., et al., *Joint computation offloading and coin loaning for blockchain-empowered mobile-edge computing.* 2019. **6**(6): p. 9934-9950.
117. Zuo, Y., et al., *Delay-limited computation offloading for MEC-assisted mobile blockchain networks.* 2021. **69**(12): p. 8569-8584.
118. Zuo, Y., S. Jin, and S.J.I.T.o.W.C. Zhang, *Computation offloading in untrusted MEC-aided mobile blockchain IoT systems.* 2021. **20**(12): p. 8333-8347.
119. Zhang, K., et al. *Mining task offloading in mobile edge computing empowered blockchain*. in *2019 IEEE International Conference on Smart Internet of Things (SmartIoT)*. 2019. IEEE.
120. Guo, S., et al., *Blockchain meets edge computing: Stackelberg game and double auction based task offloading for mobile blockchain.* 2020. **69**(5): p. 5549-5561.
121. Han, Z., et al., *Game theory in wireless and communication networks: theory, models, and applications*. 2012: Cambridge university press.
122. Chen, W., et al., *Cooperative and distributed computation offloading for blockchain-empowered industrial Internet of Things.* 2019. **6**(5): p. 8433-8446.

123. Lang, P., et al., *Cooperative computation offloading in blockchain-based vehicular edge computing networks.* 2022. **7**(3): p. 783-798.
124. Zuo, Y., et al., *Blockchain storage and computation offloading for cooperative mobile-edge computing.* 2021. **8**(11): p. 9084-9098.
125. Mitra, E.D. and W.S.J.C.o.i.s.b. Hlavacek, *Parameter estimation and uncertainty quantification for systems biology models.* 2019. **18**: p. 9-18.
126. Gu, Z.-M. and G.-G.J.F.G.C.S. Wang, *Improving NSGA-III algorithms with information feedback models for large-scale many-objective optimization.* 2020. **107**: p. 49-69.
127. Xu, X., et al., *BeCome: Blockchain-enabled computation offloading for IoT in mobile edge computing.* 2019. **16**(6): p. 4187-4195.
128. Xu, X., et al., *A blockchain-based computation offloading method for edge computing in 5G networks.* 2021. **51**(10): p. 2015-2032.
129. Xia, C., et al., *Three-stage auction scheme for computation offloading on mobile blockchain with edge computing.* 2022. **34**(25): p. e7253.
130. Li, Y., J. Wu, and L. Chen. *POEM+: Pricing longer for mobile blockchain computation offloading with edge computing*. in *2019 IEEE 21st International Conference on High Performance Computing and Communications; IEEE 17th International Conference on Smart City; IEEE 5th International Conference on Data Science and Systems (HPCC/SmartCity/DSS)*. 2019. IEEE.
131. Liu, T., et al. *Long-term auction for inner-dependent task offloading in blockchain-enabled edge computing*. in *2021 40th Chinese Control Conference (CCC)*. 2021. IEEE.
132. Seng, S., et al., *User matching on blockchain for computation offloading in ultra-dense wireless networks.* 2020. **8**(2): p. 1167-1177.